# Atmosphere expansion and mass loss of close-orbit giant exoplanets heated by stellar XUV:

## I. Modeling of hydrodynamic escape of upper atmospheric material.

I. F. Shaikhislamov[1], M. L. Khodachenko[2,3], Yu. L. Sasunov[2], H. Lammer[2], K. G. Kislyakova[2], N. V. Erkaev [4,5]

1) Institute of Laser Physics SB RAS, Novosibirsk, Russia
2) Space Research Institute, Austrian Acad. Sci., Graz, Austria
3) Skobeltsyn Institute of Nuclear Physics, Moscow State University, Moscow, Russia
4) Institute of Computational Modelling, SB RAS, Krasnoyarsk, Russia
5) Siberian Federal University, Krasnoyarsk, Russia

E-mail address: maxim.khodachenko@oeaw.ac.at

**Abstract:** In the present series of papers we propose a consistent description of the mass loss process.

To study the effects of intrinsic magnetic field of a close-orbit giant exoplanet (so-called Hot Jupiter) on the atmospheric material escape and formation of planetary inner magnetosphere in a comprehensive way, we start with a hydrodynamic model of an upper atmosphere expansion presented in this paper. While considering a simple hydrogen atmosphere model, we focus on self-consistent inclusion of the effects of radiative heating and ionization of the atmospheric gas with its consequent expansion in the outer space. Primary attention is paid to investigation of the role of specific conditions at the inner and outer boundaries of the simulation domain, under which different regimes of material escape (*free-* and *restricted- flow*) are formed. Comparative study of different processes, such as XUV heating, material ionization and recombination, $H_3^+$ cooling, adiabatic and Lyman-alpha cooling, Lyman-alpha reabsorption is performed. We confirm basic consistence of the outcomes of our modeling with the results of other hydrodynamic models of expanding planetary atmospheres. In particular, we obtain that under the typical conditions of an orbital distance 0.05 AU around a Sun-type star a Hot Jupiter plasma envelope may reach maximum temperatures up to ~9000K with a hydrodynamic escape speed ~ 9 km/s resulting in the mass loss rates ~ $(4-7) \cdot 10^{10}$ g/s. In the range of considered stellar-planetary parameters and XUV fluxes that is close to mass loss in the energy limited case. The inclusion of planetary intrinsic magnetic fields in the model is a subject of the following up paper (Paper II).

## 1. Introduction

Close orbit exoplanets in vicinity of their host stars are supposed to be exposed to intense XUV radiation which deposes significant amount of energy in the upper layers of planetary atmosphere. The fact of presence of Jupiter-type giant planets at orbital distances ≤ 0.1 AU (so called Hot Jupiters) opens questions regarding their upper atmosphere structure, its interaction with extreme stellar wind plasma flows (Kislyakova et al. 2013, 2014) and stability against escape of atmospheric gas (e.g., Guillot et al. 1996). Lammer et al. (2003) were the first who showed that a hydrogen-rich thermosphere of a Hot Jupiter at close orbital distance will be heated to several thousand Kelvin so that hydrostatic conditions will not be valid anymore and the thermosphere will dynamically expand. The stellar XUV radiation energy deposition results in heating ionization and consequent expansion of the planetary atmosphere which appear as major driving factors for the mass loss of a planet (Lammer et al. 2003, Yelle 2004, Erkaev et al. 2005, Tian et al. 2005, García Muñoz 2007, Penz et al. 2008, Guo 2011, 2013, Koskinen et al. 2010, 2013, Lammer et al., 2013). Applied hydrodynamic (HD) models by Yelle (2004), Tian et al. (2005), García Muñoz (2007), Penz et al. (2008), Lammer et al. (2009), Guo (2011, 2013) and empirical models by Koskinen et al. (2010) indicate also that close-in exoplanets experience extreme heating by stellar XUV radiation, which results in atmospheric expansion and outflow up to their Roche lobes (e.g., Erkaev et al. 2007) with mass loss rates ~$10^{10} - 10^{11}$ g s$^{-1}$. These loss rates are also supported by HST/STIS observations (Vidal-Madjar et al. 2003, 2004) which detect a 15±4% intensity drop in the high velocity part of the stellar Lyman-alpha line during the planet (HD 209458b) transit which was interpreted as a signature of neutral H atoms in the expanding planetary atmosphere.

The problem of upper atmospheric erosion of close orbit exoplanets and their mass loss is closely connected with the study of the whole complex of stellar - planetary interactions, including consideration of influences of intensive stellar radiation and plasma flows (e.g., stellar winds and CMEs) on the planetary plasma and atmosphere environments. The following effects have to be considered in that respect.

1) XUV radiation of a host star affects the energy budget of the planetary thermosphere, resulting in the heating and expansion of the upper atmosphere, which under certain conditions could be so large that the majority of light atmospheric constituents overcome the gravitational binding and escape from the planet in the form of a hydrodynamic wind (Yelle 2004, Tian et al. 2005, Penz et al. 2008, Koskinen et al. 2010, 2013, Erkaev et al. 2013, Lammer et al. 2013). This contributes to the so-called *thermal mass-loss* process of atmospheric material.

2) Simultaneously with the direct radiative heating of the upper atmosphere, the processes of ionization and recombination as well as production of energetic neutral atoms by sputtering and various photo-chemical and charge exchange reactions take place (Yelle 2004, Lammer et al. 2008, Shematovich 2012, Guo 2011, 2013). Such processes result in the formation of extended (in some cases) coronas around planets, filled with hot neutral atoms.

3) The expanding, XUV heated and photo-chemically energized, upper planetary atmospheres and hot neutral coronae may reach and even exceed the boundaries of the planetary magnetospheres. In this case they will be directly exposed to the plasma flows of the stellar wind and CMEs with the consequent loss due to ion pick-up mechanism. That contributes to the *non-thermal mass-loss* process of atmosphere (Lichtenegger et al. 2009, Khodachenko et al. 2007a). As a crucial parameter here appears the size of planetary magnetosphere, characterized by the magnetopause stand-off distance $R_S$ (Khodachenko et al. 2007a, 2007b, Lammer et al. 2009, Kislyakova et al. 2013, 2014). Altogether, this makes the planetary magnetic field and the structure of magnetosphere, as well as the parameters of the stellar wind (e.g., density $n_{sw}$ and speed $v_{sw}$) to be very important for the processes of atmospheric erosion and mass-loss of a planet.

The background magnetic field of planetary magnetosphere not only forms a barrier for the upcoming plasma flow of stellar wind, but it influences also the outflow of the escaping planetary plasma wind formed in course of atmosphere heating and ionization by stellar XUV. For example, Adams (2011) considered outflows from close-in gas giants in the regime where the flow is most likely controlled by magnetic fields. In that respect it is important to note, that the processes of material escape and planetary magnetosphere formation have to be considered jointly in a self-consistent way in their mutual relation and influence. The expanding partially ionized plasma of a Hot Jupiter atmosphere interacts with the planetary intrinsic magnetic field and appears a strong driver in formation and shaping of the planetary magnetosphere (Adams 2011, Trammell et al. 2011, Khodachenko et al. 2012), which in its turn influences the overall mass loss of a planet.

The proposed so far different modeling approaches for simulation of planetary atmosphere expansion (Yelle 2004, Erkaev et al. 2005, Tian et al. 2005, Penz et al. 2008, Guo 2011, Koskinen et al. 2013, Lammer et al. 2013), in spite of inclusion sometimes of a lot of details regarding atmospheric composition and photo-chemistry processes, quite often miss important basic processes or operate with their strong idealizations and simplifications (e.g. non-self-consistent prescription of the XUV heated region, or neglecting of gas ionization / recombination processes, etc.). Moreover, the effects of background planetary magnetic field are usually not considered at all, or again are included in a non-self-consistent way (e.g. by prescribing of specific magnetic configurations) (Trammell et al. 2011).

The investigation of exoplanetary magnetospheres by means of numerical simulations requires an efficient and well organized model capable to support a comparative study of different physical effects and processes which contribute to the formation and shaping of the magnetosphere. To investigate all the relevant processes and the role of planetary intrinsic magnetic field in the process of atmospheric material escape and mass loss, as well as formation and structuring of planetary inner magnetosphere in a self-consistent way, we adopt a *two-step modeling* strategy, starting (as a first step) with a hydrodynamic (HD) model of atmosphere expansion, which is presented in this paper. After having developed a comprehensive model for an expanding upper atmosphere of a Hot Jupiter, we will perform, as a next step, the MHD modeling, with the inclusion of planetary intrinsic magnetic fields and study of their role in formation of the inner magnetosphere and atmosphere mass loss. This part of our investigation will be reported in the following up paper.

Therefore, the major purpose of this paper is to develop a general and simple (comprehensive) enough model of an expanding atmosphere of a Hot Jupiter which includes all the major effects and yields results, consistent with the study and findings reported by other research groups (Yelle 2004, Erkaev et al. 2005, Tian et al. 2005, Penz et al. 2008, Guo 2011, Koskinen et al. 2013, Lammer et al. 2013). Stellar XUV radiation absorbed in a self-consistently defined inner layer of the modeled atmosphere is a major energy deposition source and driver for the atmosphere expansion. The effects of radiation absorption and re-emission as well as atmospheric gas ionization and recombination and gravity of planet are included in the model. Special attention in the presented numerical modelling is paid to the analysis and deeper understanding of major drivers and physical effects which result in the expansion of atmospheres of close orbit exoplanets. While considering a simple hydrogen atmosphere model, we include the effects of radiative heating and ionization of the atmospheric gas, adiabatic and radiative cooling, as well as impose more general (compared to previous studies) boundary conditions and study their influence on the simulation results. Besides of that, the influence of additional inclusion of H3+ cooling on the model behavior has been tested.

Special attention in the present study is paid to investigation of different regimes of a Hot Jupiter atmosphere expansion which are related with two basic types of 1D HD solutions, characterized by the presence of a well-formed expanding plasma wind (*free outflow* regime), or by just an insignificant slow material escape (*restricted flow* regime). In our model such distinction is achieved in a natural way by choosing different types of the outer boundary condition. As typical parameters for the modelled Hot Jupiter planet we take those of HD209458b (i.e., $R_p = 1.4 R_J, M_p = 0.7 M_J$) orbiting a Sun-like star at a distance of 0.05 AU.

For a more clear separation of various active factors influencing the formation of an exoplanetary magnetosphere, our numerical study is aimed at simulation of only the expansion of plasma, without inclusion of the rotation effects. This can be justified for the case of tidally-locked close-orbit Hot Jupiters where the rotation angular velocity of the planet is equal to that of the orbital revolution and is relatively small. In this case the radial expansion of the hot planetary plasma will dominate the co-rotation effects in the inner magnetosphere.

The paper is organized in the following way. In Section 2 we address the issue of XUV heating and radiation transfer in the upper atmosphere which play a crucial role for the modelling of a Hot Jupiter mass loss. In Section 3 the modelling concept is presented, including the atmosphere expansion drivers and role of different types of boundary conditions in that respect. Section 4 is devoted to the major equation of the presented model and key assumptions. Section 5 presents the results of the numerical modelling and related basic regimes of atmosphere expansion of a Hot Jupiter connected with specific boundary conditions applied. A possible influence of an intrinsic planetary magnetic field on the expansion flow is addressed and investigated in a number of specific cases. A comparative study of the role of different physical effects included in the model, such as XUV heating, material ionization and recombination, infrared cooling, adiabatic and Lyman-alpha cooling, Lyman-alpha reabsorption is also presented in Section 5. In Section 6 we discuss the obtained results and their relevance for Hot Jupiters.

## 2. XUV heating and radiation diffusion in upper atmosphere

A crucial issue concerns the details of the XUV heating process. The deeper is a heated layer of thermosphere, the higher energy is needed for photons to be able to reach it as the XUV absorption cross section decreases with the increasing energy of a photon as $\sigma_{XUV} \sim 1/(h\nu)^3$. An energetic photoelectron created during photo-ionization of a hydrogen atom by a photon with energy $\varepsilon = h\nu$ carries the excess energy $h\nu - E_{ion}$, where $E_{ion}$ is the ionization energy. This excess energy may be lost by the photoelectron in different ways (briefly addressed below) before it will reach a thermal balance with the entire hydrogen atoms, ions, and background electrons.

For typical conditions realized in the modeled upper thermosphere the fastest way of energy loss by a photoelectron is due to Coulomb collisions with background ions and electrons. The efficiency of heating by this process is $\eta_h \approx 100\%$. However, deeper in atmosphere (i.e., closer to the planet surface) the degree of ionization and plasma density falls down, whereas density of neutrals sharply increases. At these conditions, when the density of neutrals is higher than $\sim 10^{10}$ cm$^{-3}$, the photo-electron cools down more efficiently through inelastic collisions with atoms (while its energy exceeds excitation or ionization potential of a hydrogen atom).

Only after having lost its energy down to the values below $E_{21} \approx 10$ eV (the excitation energy), the photo-electron will start sharing the rest of its energy with the background medium through the elastic collisions with hydrogen atoms. Therefore, each absorbed XUV photon in the dense thermosphere transfers into direct thermal heating of gas a value not exceeding in average 10 eV. The excess energy is spent mostly on Lyman-alpha emission and cascading ionization. In course of the cascading ionization, the photo-electron shares with a newly released electron, in average, an energy of about $(0.2-0.5)E_{ion}$ (Kim et al. 2010). Thus, only about (0.17-0.33) of the energy $h\nu - E_{ion} >> E_{ion}$ goes into heating of overall gas in this way. At high energies ($>> E_{ion}$), the cross-section of excitation of hydrogen by electron impact and subsequent re-emission of Lyman-alpha photon is of about the same as that of ionization (see for example in Shah et al. 1987, Sweeney et al. 2001). For this energy loss channel everything depends on how optically thick is the given layer of thermosphere for Lyman-alpha photons. In the case of sufficiently high optical thickness, all the energy of photo-electrons eventually returns to the background electrons due to impact de-excitation. In opposite case it is all lost and heating efficiency reduces. Therefore, the heating efficiency $\eta_h$ of XUV may vary from about (58-67)% deep in the thermosphere down to lower values (up to ~(8-17)%) in the dense layers which are optically thin for Lyman-alpha, as well as to 100% in a fully ionized outer plasma envelope. Accordingly, a wide range of applied heating efficiencies $\eta_h$ ~(15–60) % can be found in related studies (Chassefière 1996, Yelle 2004, Lammer et al. 2009, Koskinen et al. 2013, Erkaev et al. 2013, Lammer et al. 2014). A saturation value of 93% was used in Koskinen et al. (2013). The most often used average value of the efficiency of XUV heating in the studies of exoplanet giants is $\eta_h = 50\%$. That is a value we apply in our mdelling. For an accurate calculation of the energy transport process a kinetic model of photo-electron diffusion is needed (Cecchi-Pestellini et al 2009). In that respect we note that there is an obvious distinction between the net efficiency of absorption $\eta_{net}$ which equally affects ionization and heating rates and the heating efficiency $\eta_h$ which concerns only heating.

At this point the reabsorption of Lyman-alpha emission should be also addressed. For typical temperatures of ~$10^4$ K the cross-section of Lyman-alpha photon absorption $\sigma_{L\alpha}$ by atomic hydrogen near the line center is in the range of $10^{-15} - 10^{-14}$ cm$^2$. This is much larger than the XUV absorption cross-section. Therefore, the thermosphere regions of most intense XUV heating will be optically thick for Lyman-alpha, and cooling efficiency of the last will be diminished. It is important to evaluate how strong this effect may be, as it results in higher temperatures and higher densities in the "dead-zone" of a Hot Jupiter magnetosphere and influences also the separation between "wind-" and "dead-" zones. For that purpose in Subsection 4.2 we adopt a simple model of Lyman-alpha photons spatial diffusion with a coefficient ~ $c/n_a \sigma_{L\alpha}$ as determined by distance and time intervals between consequent reabsorption events, but ignore the spectral diffusion of photon from the line center.

## 3. Modelling concept

As it was specified in the overall description of our *two-step modeling* strategy presented in Introduction, the major goal of the present work is to understand the hydrodynamic behavior of a Hot Jupiter expanding upper atmosphere without inclusion of the effects of background magnetic field of the planet. This goal may be achieved with a 1D HD model which takes into account most important physical processes which we discuss below. Even a simplified 1D HD model which deals with a one-gaseous-component (hydrogen) atmosphere and contains only the most important elements and physical processes such as heating by stellar XUV radiation, atmospheric gas ionization and recombination, as well as radiative and adiabatic cooling of the expanding planetary plasma wind, is not yet fully developed and understood due to the complex physics involved. In the present paper we propose our version of such a model.

By this, as a widely used simplification we also disregard the fact that a star illuminates only one side of a planet and assume an azimuthally symmetric radiation flux, taking into account the reduced value of $\eta_{net}$ equal to 0.5. This approach is commonly accepted as a good approximation which enables to avoid of 2-D, or extremely complex 3-D, modelling without loss of essential physics.

*3.1 Atmosphere expansion drivers*

In the first studies of the planetary atmosphere expansion aimed to estimate the material escape rate and related mass loss, the isothermal Parker-type solution was imposed with an a-priori prescribed boundary conditions at the planetary inner atmosphere (or surface) related to the XUV heating of a fixed predefined thin layer (Watson et al 1981). However, it was shown in Tian at al. (2005) that in a *thin-layer-heating* approximation the expanding atmosphere mass loss rate strongly depends on the altitude of the layer where the radiation energy is deposed.

Moreover, as it was suggested already by Parker, the realistic solution which corresponds to an outflowing (expanding) wind regime requires an additional volume heating in order to compensate the adiabatic cooling of the expanding gaseous envelope. This was also confirmed in the follow-up numerical modeling studies. For example, in the case of the solar wind, an additional acceleration factor is called for to appear at few solar radii. This additional driver is attributed to the dissipation of Alfvén waves (Usmanov et al. 2011). For an outflowing atmosphere of a Hot Jupiter one of possible candidates for the role of such a distributed accelerator is absorption of XUV, which takes place everywhere where atomic or molecular hydrogen is present. However, it turns so that at distances of about several planetary radii where the additional forcing of the expanding planetary atmospheric material is needed to keep it in the outflowing regime (i.e. to escape the gravity tension), the gas is highly ionized, and the absorption of XUV decreases.

Nevertheless, the same fact of the increasing ionization degree, which reduces the efficiency of XUV heating, contributes the additional acceleration of the wind and appears as an additional booster of the planetary plasma wind. That is because the thermal pressure $p = (n_a + n_i + n_e)kT$ increases (up to two times) at certain interval of heights (in the *ionization region*), due to the contribution of the electron and ion pressure, whereas the density of material $\approx (n_a + n_i)m_i$ remains practically unchanged. The created pressure gradient provides an additional driving force for the expansion of ionized atmospheric material. Here we use the standard notations with $n_a$, $n_i$, $n_e, m_i, k$, and $T$ staying for the number density of neutrals, ions, electrons, ion mass, Boltzmann constant, and temperature, respectively. In course of the ionization, the expanding atmospheric wind experiences in the ionization region an additional acceleration and reaches velocities which sustain the continuous outflow regime. Note, that the adiabatic cooling and the plasma pressure gradient mechanism caused by the ionization, act opposite each other, and the simple isothermal model which doesn't consider them properly produces an erroneously too strong planetary wind. To our knowledge, the role of the ionization layer and acceleration of atmospheric plasma by the pressure gradient were not elucidated so far regarding the problem of a Hot Jupiter plasma wind generation. At the same time this effect is implicitly included in several reported simulations (e.g., García Muñoz 2007, Guo 2011, 2013, Koskinen et al. 2010, 2013).

*3.2 Inner boundary conditions*

Properly defined boundary conditions at the inner edge of the simulation box appear a crucial issue for the models of atmosphere expansion driven by XUV heating. These boundary conditions are usually associated with a deep layer of inner planetary atmosphere which we conventionally call as a planet surface.

Besides of the plasma outflow and the related adiabatic cooling, due to plasma expansion, another efficient energy drain which acts to compensate the XUV heating is an excitation of the second level of atomic hydrogen by electron impact and fast emission of Lyman-alpha photon (Murray-Clay et al. 2009). At the same time, because of exponential decrease with temperature, this mechanism can fully compensate XUV heating of deep and cold layers of the thermosphere if the last decreases with height fast enough. However, due to wide spectrum of stellar XUV illuminating a planet, its heating rate decreases with height (i.e., with the increase of integrated column density) not exponentially, but more likely as a power law. Therefore, a model which is based only on the Lyman-alpha cooling requires a limiting inner boundary at the layer with a temperature of several thousand Kelvin. To admit this condition, one needs an additional justification (i.e., model) for the inner boundary at the lower thermosphere. An estimation of the overall energy budget shows that the inner "cold" thermosphere has to have a temperature of about 1000-2000 Kelvin. The cooling mechanism due to infrared emission which involves $H_2$ and $H_3^+$ species is usually addressed in that respect (Yelle 2004). Although the complex models which include the major photo-chemistry effects already exist (Yelle 2004, García Muñoz 2007, Koskinen et al 2007), in the present paper we adopt a bit different and more simple approach to take account of the infrared cooling effects.

While infrared cooling due to $H_3^+$ may dominate in the deep thermosphere at pressures larger than $\sim 1\,\mu\text{bar}$, it has been concluded in Koskinen et al. (2013), based on comparison of kinetic and hydrodynamic models, that this process doesn't affect formation of the expanding planetary wind. In that respect, Koskinen et al. (2013) used a hydrodynamic model with an inner boundary fixed at the level of $1\,\mu\text{bar}$ pressure. On the other hand, regardless of Lyman-alpha or infrared cooling, the expanding plasma motion alone, due to the adiabatic cooling mechanism, can effectively remove heat from the deep layers of thermosphere affected by XUV radiation. In some models this effect is known to be so strong that it eventually overcomes the heating and leads to an unrealistic drop in the temperature (Watson et al. 1981). However, in case when radiation (integral over spectrum) attenuates not exponentially, but more likely as inverse power of density column, the domination of adiabatic cooling effect is diminished. Below, using the results of an analytical solution which takes into account only adiabatic cooling and a proxy model of XUV heating at large integrated column densities, as well as by means of the numerical simulation, we show that the inner boundary can be extended down to the deeper layers with a much higher pressure of the order of $\widetilde{P}_0 = p_{\max} = 10\,\mu\text{bar}$. To similar conclusion arrives also García Muñoz (2007).

The question of the inner boundary pressure appears to be a crucial one also because of the fact that according to (Tian et al. 2005, Trammell et al 2011), the material escape at low pressures is practically proportional to it. This has to be avoided in a self-consistent model. Therefore, taking the inner boundary sufficiently deep in the atmosphere, where XUV energy deposition per particle becomes practically zero, seems to be a reasonable choice for a more realistic model which is independent on particular parameter values at the inner boundary, or in other words, independent on the size of the simulation box. Indeed, we found (see in subsection 5.2) that at pressures starting from $\widetilde{P}_0 = p_{sat} \sim 1\,\mu\text{bar}$ and higher, the obtained solution shows quite weak dependence on $\widetilde{P}_0$.

Another important modeling parameter related to the inner boundary pressure, that has to be properly defined is the base (or the inner boundary) temperature $\widetilde{T}_o$. General energy budget considerations constrain it to be at around $1000\,\text{K}$. The particular values of 750 K (Yelle 2004), 1200 K (García Muñoz 2007), and 1350 K (Koskinen et al. 2013) were used in the models of a Hot Jupiter. In the model presented in this paper the behavior of Hot Jupiter environment insignificantly depends on particular value of $\widetilde{T}_o$. That is because of taking the inner boundary of the simulation box at sufficiently deep layers with high enough pressure which are unaffected by the stellar XUV. At such deep located inner boundary an extremely steep (exponential) density grow takes place, so that the condition of a constant mass flux supposes very small values for the velocity. Therefore, taking velocity equal zero at the inner boundary does not affect the numerical solution at all, if the boundary is sufficiently deep.

Inclusion in the model of deep thermosphere layers with pressures higher than $\sim 1\,\mu\text{bar}$, requires certain attention to the infrared cooling effect due to $H_3^+$ and understanding of how it affects the simulation results. To check this we performed calculations with a proxy model for the infrared cooling with volume rates about $10^{-7}\,\text{erg}\,\text{cm}^3\,\text{s}^{-1}$ reported in (Yelle 2004, García Muñoz 2007). The obtained results enable us to conclude that temperature profile in the inner atmosphere regions (i.e., close to the planet surface) and the corresponding density distribution are somewhat sensitive to the additional energy loss mechanism related with the infrared cooling (besides of the adiabatic and Lyman-alpha cooling). But due to its local character and action first of all in deep layers of the inner thermosphere, the $H_3^+$ cooling leads to the reduction of the simulated material escape rate approximately only by a factor of two. Such insignificant difference gives a reason for an essential simplification of the model of atmosphere expansion and mass loss by ignoring complex hydrogen chemistry at the base of cold thermosphere, or use of its simplified modeling approximations.

*3.3 Outer boundary conditions*

An appropriate definition of conditions at the outer boundary also plays an important role for the modeling. Similarly to the case of inner boundary conditions discussed above, it would be reasonable to have outer boundary conditions also defined so, that the obtained numerical solutions are quantitatively independent on them in a sufficiently wide range of parameters.

Besides of that, it is necessary to distinguish between two different types of outer boundary conditions that are related to different external physical circumstances. One is an *open* (or free) outer boundary, when the expanding atmosphere material can leave the vicinity of a planet, i.e. when the escape of mass and energy takes place, and the escaping plasma can flow through the outer boundary unaffected. Another situation corresponds to the *closed* (or shut-up) conditions when the material flow is *suppressed* or *restricted* at certain distance from a planet by some external physical factors not included self-consistently in the model. As an example of such restricting factors, an encountering stellar wind or planetary magnetic field can be referred. The choice of particular type of outer boundary conditions finally depends on a goal of specific modeling and a region of application of the modeling results. In particular, in the case of a Hot Jupiter, modeling of different parts of the magnetized planetary extended thermosphere may require different sets of the external boundary conditions.

For example, simulation of a so-called "wind zone" of a Hot Jupiter magnetosphere (Adams 2011, Trammell et al. 2011, Khodachenko et al. 2012) at sufficiently high latitudes where plasma leaves the planet along the open magnetic field lines needs the *open* outer boundary conditions, whereas modeling of the "dead zone" occupied by a stagnated plasma of inner magnetosphere locked in the region of closed magnetic field lines would requires the *closed* boundary.

Besides of that, the interaction of the magnetized Hot Jupiter with a stellar wind results in the formation of a magnetosphere obstacle located at several planetary radii from a planet. Depending on a character of the encountering stellar wind plasma flow (sub-, or super- sonic/Alfvénic) this magnetosphere obstacle can have a form of a paraboloidal shock or Alfvén wing system (Ip et al. 2004, Erkaev et al. 2005, Khodachenko et al. 2012) elongated in the direction of the plasma flow, which for the close-orbit planets has to take into account also the Keplerian velocity of orbital motion of the planet (Grießmeier et al. 2007, Cohen et al. 2011, Khodachenko et al. 2012). Namely because of these factors, the outer boundary conditions in the first numerical modelling of a hydrodynamic atmosphere escape on a Hot Jupiter (Yelle 2004) were imposed at a relatively close distance of only $3R_p$ from the planet. Note, that a question regarding the interaction of an upcoming stellar wind with the expanding planetary plasma wind, and the role of planetary magnetic field in this process was not considered in Yelle (2004) as well as in the later modeling cases (García Muñoz 2007, Penz et al. 2008, Guo 2011, 2013, Adams 2011, Trammell et al. 2011, Koskinen et al. 2013, Khodachenko et al. 2012, Lammer et al. 2013, Erkaev et al. 2013). Thus it still remains a subject for future investigation.

# 4. Model equations

*4.1 Basic equations and major assumptions*

In this section we describe the basic equations of the presented numerical model. For simplicity's sake we restrict the modeling by assuming a purely hydrogen atmosphere of a Hot Jupiter composed in general case of hydrogen atoms, ions and electrons. As it has been shown in García Muñoz (2007), Trammell et al. (2011), Koskinen et al. (2010), and Guo (2011), the atomic and Coulomb collisions in the modeled regions of a Hot Jupiter upper atmosphere are efficient enough to ensure the "non-slippage" approximation for neutral hydrogen atoms, ions, and electrons, so that the assumptions of $\mathbf{V}_a = \mathbf{V}_i = \mathbf{V}_e = \mathbf{V}$ as well as temperature equilibrium $T_a = T_i = T_e = T$ hold true. Here $\mathbf{V}_k$ and $T_k$ with $k = a, i, e$ correspond to the velocity and temperature of components of the partially ionized atmospheric gas, respectively; whereas $\mathbf{V}$ and $T$ stay for its velocity and temperature as a whole.

For the typical temperatures $\leq 10^4$ K realized in the upper atmosphere of a Hot Jupiter (Koskinen et al. 2010), the proton-proton Coulomb collision cross-section is well above $10^{-13}$ cm$^{-2}$. The proton-hydrogen collisions are dominated by resonant charge-exchange with cross-section, at the typical velocity of 10 km/c, approximately $10^{-14}$ cm$^{-2}$. The location of exobase for a Hot Jupiter which corresponds the region where Knudsen number $Kn = \bar{l}/\Delta$ reaches 1 is determined by the condition when the mean free path $\bar{l}$ is equal to a typical scale of flow $\Delta$. In the regions far from the planet $\Delta$ may be taken to be of the order of the planetary radius, i.e. $\Delta \sim R_p$, and the condition $Kn \sim 1$ is realized approximately in the layers with density $\sim 10^4$ cm$^{-3}$. That remains well above the height range ($10 R_p$) considered in the modelling simulations. Close to the planet surface, the barometric height $H$, which is smaller than $R_p$, has to be taken as the typical scale $\Delta$. However, due to much larger densities there, the mean free path $\bar{l}$ also becomes smaller and the condition $Kn < 1$ holds true. Therefore, the fluid approach adopted in the present study is fairly valid.

Within the frame of made assumptions, taking into account the processes of hydrogen ionization (by XUV and electron collision) and recombination, the continuity, momentum, and energy balance equations of the model can be written as follows:

$$\frac{\partial}{\partial t} n + \nabla(\mathbf{V} n) = 0 \tag{1}$$

$$\frac{\partial}{\partial t} n_i + \nabla(\mathbf{V} n_i) = n_a \langle \sigma_{XUV} F_{XUV} \rangle - n_i \Lambda_e \sigma_{rec} + n_a \Lambda_e \sigma_{ion} \tag{2}$$

$$m \frac{\partial}{\partial t} \mathbf{V} + m(\mathbf{V}\nabla)\mathbf{V} = -\frac{1}{n}\nabla nkT - \frac{1}{n}\nabla n_e kT + m\nabla \frac{GM_p}{r} \tag{3}$$

$$\frac{3}{2}(n + n_e)k \cdot \left( \frac{\partial}{\partial t}T + (\mathbf{V}\nabla)T + (\gamma - 1)T \nabla \mathbf{V} \right) = n_a \langle (E_{XUV} - E_{ion}) \sigma_{XUV} F_{XUV} \rangle$$
$$- E_{21} \Lambda_e \cdot (n_a \sigma_{12} - n_{a, n=2} \cdot \sigma_{21}) - n_a E_{ion} \Lambda_e \sigma_{ion} \tag{4}$$

where the following notations are used:

$$n = n_a + n_i, \; n_e = n_i = x_{ion} n \tag{5}$$
$$\Lambda_e = n_e V_{Te}$$

In the equations (1)-(4) we use the standard cross-sections of major processes, such as ionization by XUV: $\sigma_{XUV} = 6.3 \cdot 10^{-18}$ cm$^2$; ionization by electron impact: $\sigma_{ion} = 4 \cdot 10^{-16} \cdot e^{-E_{ion}/T} \cdot T^{-1}$ cm$^2$; recombination with an electron: $\sigma_{rec} = 6.7 \cdot 10^{-21} T^{-3/2}$ cm$^2$; hydrogen excitation and de-excitation: $\sigma_{12} = \sigma_{21} \cdot e^{-E_{21}/T}$ cm$^2$ and $\sigma_{21} = 7 \cdot 10^{-16} T^{-1}$ cm$^2$, respectively; Lyman-alpha absorption: $\sigma_{L\alpha} \sim (10^{-15}...10^{-14}) T^{-1/2}$ cm$^2$ with the temperatures scaled in units of characteristic temperature of the problem ($\widetilde{T}_c = 10^4$ K), except of exponent power indices in the expressions for $\sigma_{ion}$ and $\sigma_{12}$, where temperature is taken in energy units. Besides of that, for the wavelength averaged terms corresponding to XUV ionization and heating rates in equations (2) and (4) we use the proxy model proposed in Trammell et al. (2011):

$$\langle \sigma_{XUV} F_{XUV} \rangle = \frac{t_{XUV}^{-1}}{1 + (\sigma_{XUV} NL)^{1.5}} \qquad (6)$$

$$\langle (E_{XUV} - E_{ion}) \sigma_{XUV} F_{XUV} \rangle = \frac{\varepsilon_{XUV} t_{XUV}^{-1}}{(1 + 0.2 \cdot \sigma_{XUV} NL)^{1.2}}, \qquad (7)$$

where $NL = \int_r^\infty n_a dr$ is the neutral hydrogen column density.

The typical values for the mean ionization time $t_{XUV} \approx 6$ hour and mean photoelectron energy (or energy per absorption event) $\varepsilon_{XUV} \approx 2.7$ eV, that goes to heating, are taken in (6) and (7) to correspond the integrated stellar XUV energy flux of $F_{XUV} \approx 2 \cdot 10^3$ erg cm$^{-2}$ s$^{-1}$ expected at 0.05 AU orbit around a Sun-analogue star. Note that the decrease of heating (equation (7)) is close to an inverse rather than exponent power of density column. The reasons to use the proxy model instead of one calculated with the frequency binned real Solar spectra are as follows. The deviation of proxy model from the spectrum-based one is sufficiently small to ensure that error in computed mass-loss rate is less than 25%. Besides of that, using the analytical model enables to make simple analytical estimations of the expected solutions which help to verify the correctness of the obtained numerical results (see in Sect. 5.2) and to understand physics of the involved processes.

To take into account the fact that the star illuminates only one side of the planet, in our calculations we reduce the radiation flux (and the corresponding values given by (6) and (7)) by a factor of two, i.e. take $\eta_{net} = 50\%$. Besides of that, for definiteness' sake we suppose the heating efficiency $\eta_h$ of stellar XUV (Chassefière 1996, Yelle 2004, Lammer et al. 2009, Koskinen et al. 2013) to be 50 % and therefore divide by two once more the outcome of (7) which is then used as an approximation for the term $\langle (E_{XUV} - E_{ion}) \sigma_{XUV} F_{XUV} \rangle$ in the energy equation (4).

*4.2 Energy balance and radiation diffusion equations*

Besides of the XUV heating term and adiabatic cooling due to plasma expansion (the last term at the left hand side of eq. (4)), the energy equation (4) takes into account also energy loss and gain due to excitation and de-excitation of $n = 2$ level of hydrogen atom by electron impact, as well as cooling due to ionization. We assume that the concentration of excited atoms remains sufficiently low, i.e. $n_{a,n=2} \ll n_{a,n=1} \approx n_a$. The validity of this assumption is specially verified in course of the modeling simulation. Because of a too low rate at considered densities, we also neglect in equation (2) the process of three-body recombination which is reverse to the inelastic impact ionization.

To complete the energy equation (4) one has to define a way for calculation of concentration $n_{a,n=2}$ of the excited atoms at the level $n=2$. In our model it is controlled by density of Lyman-alpha photons which are emitted by the excited atoms and re-absorbed by the atoms at non-excited state (i.e. level $n=1$). In case of an optically thin gas envelope, the concentration of excited atoms remains very small (negligible), due to fast radiative de-excitation. On the other hand, in the optically thick case, the concentration $n_{a,n=2}$ may remain moderately high due to Lyman-alpha photons sequential re-absorptions until the photons leave the system. Such a process of multiple re-absorptions and re-emissions of a photon with a chaotic change of its travel direction (after each re-emission act) can be described in the first approximation as diffusion with a coefficient determined by the average distance $(n_a \sigma_{L\alpha})^{-1}$ and velocity $c$ between re-absorptions. The corresponding diffusion equation for the Lyman-alpha photons density $N_{La}$ and concentration of excited atoms may be written as follows:

$$\frac{d}{dt} n_{a,n=2} = \Lambda_e \cdot (n_{a,n=1} \sigma_{12} - n_{a,n=2} \sigma_{21}) - \frac{n_{a,n=1}}{\tau_{21}} + n_{a,n=1} \cdot c N_{La} \sigma_{La} \qquad (8)$$

$$\frac{\partial}{\partial t} N_{La} = \frac{n_{a,n=2}}{\tau_{21}} - n_{a,n=1} \cdot c N_{La} \sigma_{La} + div \frac{c}{n_a \sigma_{La}} \nabla N_{La} \qquad (9)$$

The terms in the right-hand side of equation (9) stay for the photon emission by the excited atoms with mean emission time $\tau_{21}$, absorption by non-excited atoms, and spatial diffusion, respectively. For a typical scale of system being about of a radius of planet $R_p \sim 10^{10}$ cm and $\sigma_{La} \sim 10^{-15}$ cm$^2$, the gas becomes optically thick for Lyman-alpha photons at densities above $10^5$ cm$^{-3}$.

Introducing a normalized number density of Lyman-alpha photons $\tilde{N}_{L\alpha} = \tau_{21} c \cdot N_{L\alpha} \sigma_{L\alpha}$, the stationary concentration of the excited atoms may be expressed from equation (8) as:

$$n_{a,n=2} = n_{a,n=1} \frac{\tilde{N}_{L\alpha} + \Lambda_e \cdot \sigma_{12} \tau_{21}}{1 + \Lambda_e \cdot \sigma_{21} \tau_{21}} \qquad (10)$$

By substitution of this solution in equations (4) and (9), and taking into account that $n_{a,n=1} \approx n_a$ we obtain a modified energy balance equation and the equation for normalized number density of Lyman-alpha photons:

$$\frac{3}{2}(n+n_e)k \cdot \left(\frac{\partial}{\partial t}T + (\mathbf{V}\nabla)T + (\gamma-1)T \nabla \mathbf{V}\right) = n_a \langle (E_{XUV} - E_{ion}) \sigma_{XUV} F_{XUV} \rangle \qquad (11)$$
$$- n_a E_{21} \Lambda_e \sigma_{21} \cdot (e^{-E_{21}/T} - \tilde{N}_\alpha) - n_a E_{ion} \Lambda_e \sigma_{ion}$$

$$\frac{\partial}{\partial t} \tilde{N}_{L\alpha} = \tau_{21} c \cdot n_a \sigma_{L\alpha} \cdot \Lambda_e \cdot \sigma_{21} (e^{-E_{21}/T} - \tilde{N}_{L\alpha}) + \nabla \cdot \left(\frac{c}{n_a \sigma_{L\alpha}} \nabla \tilde{N}_{L\alpha}\right) = 0 \qquad (12)$$

Since the emission time $\tau_{21} \sim 1\,ns$ is very small, we neglect the terms proportional to $\Lambda_e \cdot \sigma_{12} \tau_{21}$ and $\Lambda_e \cdot \sigma_{21} \tau_{21}$ in the final form of the equations (11) and (12), as compared with unit.

When the system is essentially optically thick, the diffusion term in (12) tends to zero and the number density of Lyman-alpha photons saturates at a level where collisional excitation and de-excitation balance each other, resulting in $\widetilde{N}_{L\alpha} = e^{-E_{21}/T}$. Note that for temperatures $T \ll E_{21}$ the population of excited atoms is much smaller than that of non-excited, ground state atoms: $n_{a,n=2}/n_a \approx \widetilde{N}_{L\alpha} \ll 1$. In that respect, one of the reasons for inclusion in the equation (11) of a relatively unimportant electron impact ionization term is to ensure a non-zero cooling in case of total saturation of $n = 2$ level by Lyman-alpha photons.

Further on, we represent temperature and velocity of the simulated upper atmospheric material in units which correspond to the characteristic temperature of the problem taken as $\widetilde{T}_c = 10^4$ K and the corresponding $\widetilde{V}_c = V_{Ti} = 9,8$ km/s, whereas the distances will be normalized to a typical scale of the system, defined by planetary radius $R_p$.

## 5. Simulation results

The numerical solving of the model equations (1)-(3), (11) and (12) was performed with an explicit second order differential scheme on a non-uniform grid with a mesh size taken proportional to the radial distance to resolve the decreasing barometric scale height near the planet surface. The dimensionless mesh size (normalized to $R_p$) at the inner boundary was taken (depending on particular run) as $\Delta r_{min} = (1 - 2.5) \cdot 10^{-3}$. The time step was constant and of the order of $\Delta t \leq 10^{-1} \Delta r_{min}$. Discretization is central based. Time marching is a simple one stage explicit iteration. Stationary solution is sought with convergence criteria defined as the relative change of calculated values (for example, of the mass-loss) at the outer boundary at one step not exceeding $10^{-4}$. Usually it takes several hundred to several thousand dimensionless times or $10^5 - 10^6$ iterations to achieve the steady state. The spherical geometry with all physical values varying only along radial distance was considered. Because of large density gradients realized in course of the simulation, to increase the accuracy of calculations, instead of real density values in the continuity equation (1) we dealt with $\ln(n)$. In order to suppress instability of the numerical scheme related with generation of acoustic-gravity waves, a small artificial viscosity term $\nabla \eta \nabla \mathbf{V}$ was added in the momentum equation (3) with a numerical coefficient $\eta$ about $10^{-5}$ in dimensionless units.

The initial profile of neutral thermosphere density in hydrostatic and isothermal equilibrium is given by the solution of equation (3) in steady state:

$$n_{ao}[cm^{-3}] = \frac{10^{13}}{1.38 \cdot T_o} \exp\left(-\frac{\lambda}{T_0}(1 - 1/r)\right), \quad (13)$$

where $\lambda = GmM_p/(R_p k \widetilde{T}_c)$ is an important dimensionless parameter of the system defined as the relation of thermal energy to gravitation escape energy at the inner boundary of the simulation region, $r = 1$ (in units of $R_p$), i.e. close to the conventional planet surface. For the considered parameters of a Hot Jupiter, $\lambda$ has a typical value of about 10. The coefficient before exponent in equation (13) is taken so, that the density value at the inner boundary of the simulation region corresponds, for the given temperature, to the pressure of $\widetilde{P}_0 = 10^{-5}$ bar. At the same time, from the computing point of view, the exact position of this boundary (at the surface of a planet, or slightly above – in the deep atmosphere) is not important because of very steep dependence of density on $r$ at the initial dimensionless temperature $T_o = 0.1$ which corresponds an absolute inner boundary temperature value of $\tilde{T}_0 = 1000\,\text{K}$.

*5.1 Two basic regimes of the atmosphere expansion*

Below we consider numerical solutions of a spherical 1D HD model which describe different behavior regimes of a Hot Jupiter's upper atmosphere, corresponding to the two basic types (i.e., *open* and *closed*) of the outer boundary conditions.

The position of outer boundary is taken so far that the results of calculations, such as mass loss rate do not depend on the distance of its particular location. The question discussed in the literature is the position of exobase relative to the model outer boundary. For HD modeling of some particular Hot Jupiters (e.g., HD209458b) it was shown in several papers that the exobase is farther than $10 R_p$. Thus in our modeling we also take the location of outer boundary at $10 R_p$, where the numerical solution practically doesn't depend on the boundary exact position and use of hydrodynamic modeling approach is justified. More detailed (García Muñoz 2007) or partially kinetic conditions (Koskinen et al. 2013) also lead to similar results and conclusions.

When the cooling mechanism due to Lyman-alpha radiation is taken into consideration, the model admits two general solutions related to the global energy balance. One of the solutions represents the already mentioned above case with an atmospheric material *free outflow* when the XUV heating is balanced by the advection and adiabatic cooling. Another solution corresponds to a *restricted-flow* regime, in which the XUV heating is balanced mostly by Lyman-alpha cooling. The particular values of physical quantities are also different for these two types of the solutions. By this, the *restricted-flow* solution is characterized by a hotter and about an order of magnitude denser stagnated plasma envelope. An important issue regarding the *free outflow* and the *restricted-flow* solutions consists in the fact, that their realization in our model is completely controlled by specific outer boundary conditions only. Note, that the *free outflow* and *restricted-flow* regimes are physically relevant (and may be attributed) for the "wind-" and "dead-" zones of a Hot Jupiter magnetosphere, respectively (Trammell et al. 2011, Khodachenko et al. 2012).

As it has been mentioned above (see Subsection 3.2), the Lyman-alpha cooling is inefficient at temperatures below several thousand Kelvin. Therefore, a small outflow of material is needed to compensate the excessing XUV heating in such a dense deep thermosphere. In that respect, instead of a totally closed (shut-up) condition at the outer boundary we allow a small flow of material at a speed of the order of $0.1\,\text{km/s}$. This in fact constitutes the matter of the *restricted-flow* regime. Such an assumption is relevant for the conditions in a "dead-zone" of Hot Jupiter's inner magnetospheres or in the front region of interaction with a stellar wind, as a small portion of the planetary plasma might still leave the planet there by convection to higher latitudes or to the tail.

To deal with the *open* outer boundary in practice, we assume spatial derivatives of physical values to be small, i.e. asymptotically approaching to zero with a decreasing mesh size and increasing distance.

Similar approach to the study of atmosphere expansion regimes has been adopted for the first time in (García Muñoz 2007) where instead of velocity, a fixed pressure was imposed at the outer boundary, and the solutions with an escaping material flow varying from a super-sonic up to slow subsonic one were obtained.

*5.2 Free outflow regime of a Hot Jupiter's atmosphere expansion*

Figure 1 shows the result of modeling calculation with typical model parameters summarized in the Table 1. The case of an *open* outer boundary condition was considered. In line with the discussion in Subsection 3.2, the Lyman-alpha re-absorption effects are not important for the studied regime of material *free outflow* and therefore, were not included in the model.

**Table 1.** Model parameters.

| $M_p$ | $R_p$ | $F_{XUV}$ (i.e., 0.05 AU orbit, Sun-analogue star) | $\widetilde{T}_0$ | $\widetilde{P}_0$ | Outer boundary condition |
|---|---|---|---|---|---|
| $0.7 M_J$ | $10^{10}$ cm | $2 \cdot 10^3$ erg cm$^{-2}$ s$^{-1}$ | 1000 K | $10^{-5}$ bar | *Open* |

As it can be seen in Figure 1, the maximum temperature realized in the *free outflow* regime is about 8500 K and is achieved at a height of about $1.75 \cdot R_p$. Then, due to the adiabatic cooling the temperature quickly decreases with height because the XUV heating is not enough to compensate it once the ionization of the expanding gas takes place. The ionization degree $x_{ion} = \frac{n_i}{n}$ increases sharply near the temperature maximum, and the velocity of the expanding atmospheric plasma wind driven by the increase of pressure gradient, also increases mostly at the ionization front. Since the temperature and density quickly decrease with distance, the escaping material reaches supersonic velocity at a distance of about $6 \cdot R_p$.

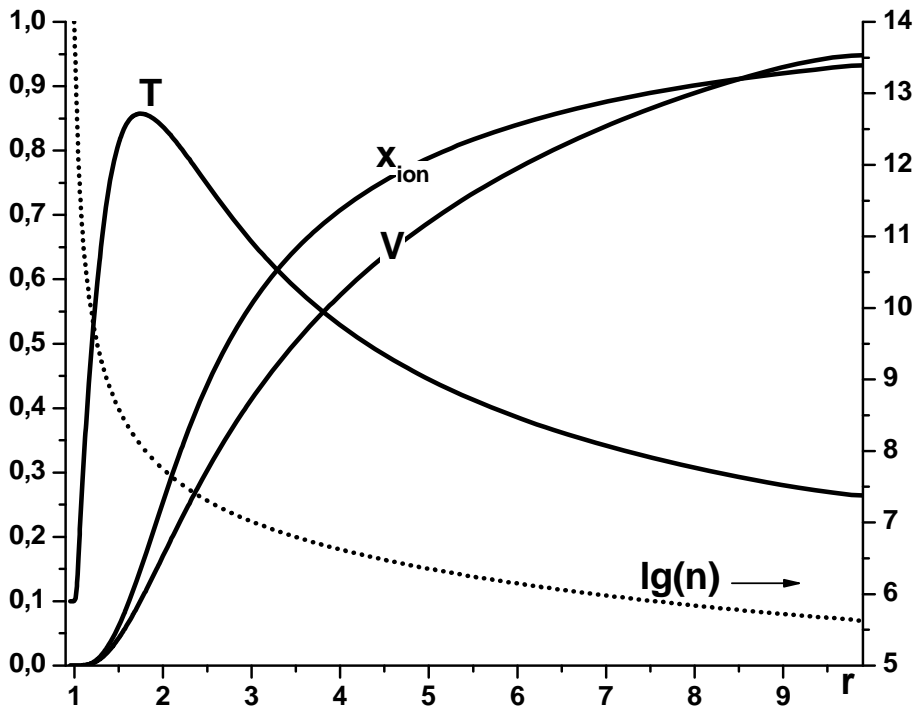

**Figure 1.** Profile of temperature, velocity, ionization degree (left axis) and density (dotted line, right axis, log scale) in dimensionless units obtained in calculation with typical model parameters from Table 1.

The comparison of obtained results with outcomes of the closest (in the sense of model parameters and simulation conditions) model by Koskinen et al. (2013) reveals that in spite of general (phenomenological) agreement, the models show some non-principal difference in specific physical quantities. These may be attributed to slightly different values of XUV flux used in the modeling simulations, which result in different heating rates $q_{XUV} = n_a \langle (E_{XUV} - E_{ion}) \sigma_{XUV} F_{XUV} \rangle$ in the equation (4) (and (11)) realized in course of modeling, and to different net and heating efficiencies. In particular, our model gives slightly higher density and a bit smaller velocity with a slightly smaller (than by Koskinen et al. (2013)) maximum XUV heating rate: $q_{XUV}^{max} \approx 4 \cdot 10^{-8}$ erg·cm$^{-3}$·s$^{-1}$.

The calculation of mass loss rate $\dot{M} = 4\pi mnr^2 V_r$ yields the value $7 \cdot 10^{10}$ g/s, which is by a factor of almost two larger than that predicted with the model by Koskinen et al. (2013). However, the mass loss rates become practically equal if we take the same net efficiency $\eta_{net} = 25\%$ as that used by Koskinen et al. (2013).

The most significant difference of our model results from those of Koskinen et al. (2013) concerns the predicted maximum temperature, which in our case is by 3500 K less. One of the reasons is that Koskinen et al. (2013) use the reduced by a factor of 0.1 Lyman-alpha cooling. Switching-off the Lyman-alpha cooling in our model reduces the difference down to 2500 K.

Regardless of the maximum temperature, the value of pressure averaged temperature

$$\bar{T}_p = \frac{\int_{p_1}^{p_2} T(p) \mathrm{d}(\ln p)}{\ln(p_2/p_1)} \qquad (14)$$

over the distance interval from $R_p$ to $3R_p$, which is used in Koskinen et al. (2013) as a measure of possible transit effect of a planet, in the case of our model is 6000 K. That is rather close to the temperature range of (6200 – 7800) K, reported in Koskinen et al. (2013).

The behavior of the temperature and density of the expanding planetary atmosphere as parametric functions of pressure in the deep region of dense thermosphere layers near the inner boundary, obtained by solving numerically the model equations, is shown in Figure 2. The position of the inner boundary itself in this study was taken slightly (a few percent) below the level of $r = 1$ (in units of $R_p$). Note that temperature smoothly approaches the base (inner boundary) temperature $T_o = 0.1$ (in units of $\tilde{T}_c = 10^4 K$) while velocity decreases down to negligibly small values.

To understand the nature of the obtained numerical result let's compare it with a simple analytical solution. Close to the inner boundary, in the dense layers of thermosphere where velocity is rather small, we can ignore inertial terms and ionization in the momentum equation (3). Then, barometric density profile $n(r) = n_0 \exp(-r/H(T,r))$ with the local scale height $H(T,r) = r^2 kT/(GmM_p)$ is an approximate solution of equation (3), which under the above mentioned conditions has the following form:

$$\frac{k}{n} \nabla nT \approx -\nabla \frac{GmM_p}{r} \qquad (15)$$

We assume that XUV absorption takes place in a thin layer near $r \approx R_p$ with very large column densities $NL$, so that $\sigma_{XUV} NL \approx \sigma_{XUV} nH \gg 1$; and it is balanced by adiabatic expansion.

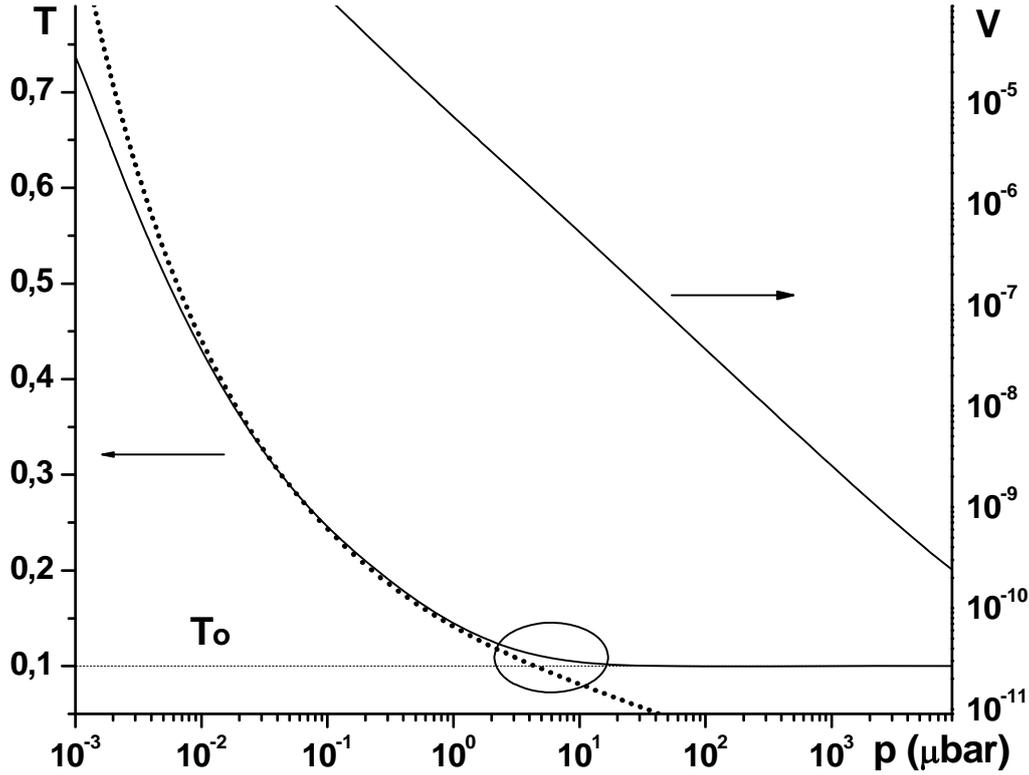

**Figure 2.** Temperature and velocity in dimensionless units as functions of pressure at the inner boundary in the thermosphere. Dotted line – analytic solution defined with equations (16) and (17).

Then, from the energy balance equation (4) (or (11)) in a quasi-stationary state, using the approximation (7) taken at $NL \gg 1$, assuming as a zero-order approximation the barometric density distribution with the scale height $H(T,r)$, and expressing $r^2 V_r$ in terms of mass loss rate $4\pi mnr^2 V_r = \dot{M} = const$, we obtain an approximate relation between temperature and density in vicinity of the inner boundary:

$$\left(\frac{T}{T_o}\right)^6 \left(\frac{R_p}{r}\right)^8 \frac{n}{n_o} = (\sigma_{XUV} H_o n_o)^{-1} \left(\frac{4\pi R_p^2 mEF}{\sigma_{XUV} k \widetilde{T}_o \dot{M}}\right)^5 = const, \qquad (16)$$

where $EF = \varepsilon_{XUV}/(0.2^{1.2} \cdot t_{XUV})$, $T_o$ is the inner boundary temperature measured in units of $\widetilde{T}_c = 10^4 K$, and $H_o = H(\widetilde{T}_0, R_p)$ is the local scale height in vicinity of the inner boundary. In these calculations we continue using $t_{XUV} \approx 6$ hour and $\varepsilon_{XUV} \approx 2.7$ eV, which correspond the stellar XUV flux expected at 0.05 AU orbit around a Sun-analogue star.

Solution of barometric equation (15), with the expression (16) taken into account, yields

$$\frac{T}{T_o} = (r/r_{min})^{8/5} + \frac{1}{13 H_o r}\left[(r/r_{min})^{13/5} - 1\right]. \qquad (17)$$

The starting point $r_{min}$ of the solution is determined from (16) by taking $T = T_o$ and $n = n_o$. The maximum pressure (or density, i.e. $p_{max} / k\widetilde{T}_0$) up to which solution (17) is valid, follows from the right hand side of (16) with $const = 1$:

$$p_{max} = n_0 k\widetilde{T}_0 \approx \frac{k\widetilde{T}_o}{\sigma_{XUV} H_o} \left( \frac{4\pi R_p^2 mEF}{\sigma_{XUV} k\widetilde{T}_o \dot{M}} \right)^5 \tag{18}$$

The term in brackets in equation (18) is approximately a ratio of energy delivered by XUV to energy removed by the expanding material flow, if each particle has energy $k\widetilde{T}_o$. Because the material escape is almost energy limited and particle energy is ten times higher than $k\widetilde{T}_o$, the term in brackets is about $20-30$. It means that maximum pressure is much larger than that at the layer with a density of $1/\sigma_{XUV} H_O$ where a significant portion of XUV is absorbed. Typical Hot Jupiter conditions in this case yield an estimate for pressure: $p_{max} \sim 10\,\mu bar$.

The analytic solution (17) expressed by means of equation (16) as a function of pressure is shown in Figure 2. By this, the value $p_{max} = 10\,\mu bar$ well corresponds to the point (marked by an ellipse) where the numerical solution, approaching the inner boundary level, begins to deviate significantly from the analytical solution (16). Note that in this region the velocity is already very small. At larger pressures, corresponding to deeper atmospheric layers, the considered model solution is governed more by numerical factors such as small viscosity ($\eta = 10^{-5}$ in dimensionless units) introduced into the equation (3) for stability reasons. At such very large pressures and so deep in the thermosphere, the hydrogen chemistry, molecular and eddy diffusion are probably more important than the adiabatic cooling considered in this example.

*5.3 Planetary atmosphere mass loss: influence of the inner boundary position and XUV flux*

To investigate how the inner boundary location may influence the obtained numerical results regarding the escape of Hot Jupiter atmospheric material and to verify a possibility to neglect the effects of infrared cooling by $H_3^+$ at the base of thermosphere (presently assumed), we performed a series of dedicated calculations, and present their outcome below.

Figure 3 shows a dependence of the simulated mass loss rate on a value of fixed pressure $\widetilde{P}_o$ at the inner boundary, or rather on the position of the inner boundary relative to fixed distribution (13). One can see that for $\widetilde{P}_o$ less than $p_{sat} \sim 1\,\mu bar$ the escaping flow is strongly affected by a particular value of the inner boundary pressure $\widetilde{P}_o$. Such a dependence of the numerical solution on the inner boundary condition (i.e. on the location of the inner boundary of the simulation box) has been referred in several previous studies (e.g., Trammell et al. (2011), Adams (2011)), where proportionality of the mass loss rate to the boundary pressure was assumed as an input condition. As one can see from Figure 3, for $\widetilde{P}_o$ higher than $p_{sat} \sim 1\,\mu bar$, the influence of the inner boundary pressure on mass loss rate become to be less significant.

As to the behavior of other physical quantities in the model, we find that between the cases of $\widetilde{P}_0 = 100\,\mu bar$ and $\widetilde{P}_0 = 0.1\,\mu bar$ the material density at distances of few $R_p$ varies by a factor of two, whereas the location of temperature maximum for $\widetilde{P}_0 = 100\,\mu bar$ shifts by $\sim 0.3 R_p$ to smaller heights.

The choice of particular maximum pressure at the inner boundary is also related to the depths of absorption of XUV photons. The absorption parameter $\sigma(\varepsilon)\cdot NL$ for a photon of given energy $\varepsilon = h\nu$ can be estimated as:

$$\sigma(\varepsilon)\cdot NL \approx \sigma_{XUV}(E_{ion}/\varepsilon)^3 n_o H_0 \approx \sigma_{XUV}(E_{ion}/\varepsilon)^3 R_p^2/(GmM_p)\cdot \widetilde{P}_o \approx 0.5\cdot 10^{10}(E_{ion}/\varepsilon)^3 \widetilde{P}_o$$

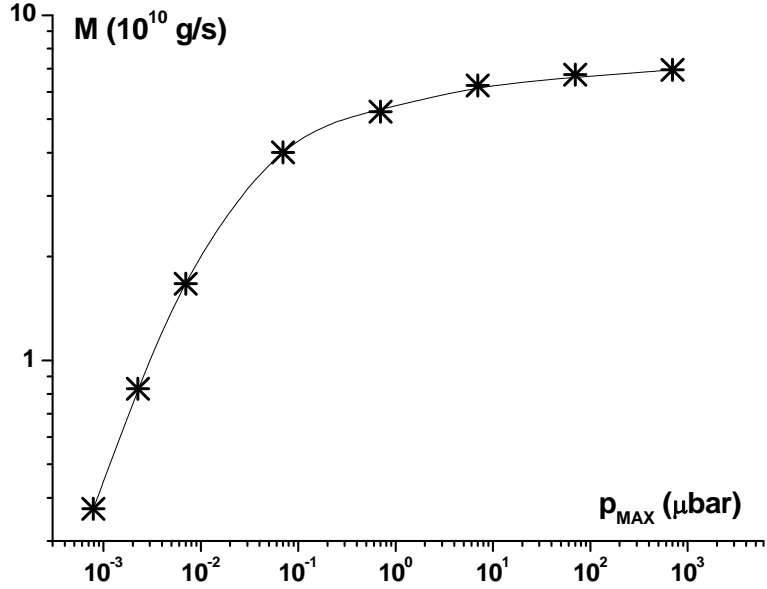

**Figure 3.** Mass loss rate as a function of the inner boundary pressure $\widetilde{P}_0$.

Therefore, fixing of the thermosphere inner boundary (i.e., depth) at the level of $\widetilde{P}_o = 1\,\mu\mathrm{bar}$ means that photons with energy higher than $\varepsilon > 20\cdot E_{ion}$ which correspond approximately to wavelengths shorter than 5 nm will be excluded from the consideration. Such a restriction would limit the generality of the model.

The discovered mass loss "saturation" effect for sufficiently large inner boundary pressures ($\widetilde{P}_o > 1\,\mu\mathrm{bar}$) is related with a rather weak (for large $\widetilde{P}_o$) dependence of the obtained numerical solution on the base temperature $\widetilde{T}_o$. In particular, variation of $\widetilde{T}_o$ from 750K to 1500K results in change of mass loss rate $\dot{M}$ by only 14%.

Figure 4 shows the dependence of the mass-loss rate on intensity of XUV radiation of a Sun-like star illuminating the planet (full analog of HD209458b), expressed in terms of the orbital distance to the star. It has a power-law character. The fact that the increase of the mass-loss rate with decreasing distance is very close to $R^{-2}$ enables to conclude on the proportionality between the mass loss rate and the XUV flux. Similar behavior has been reported in a number of papers (e.g., Lammer et al. 2003, Garcia Munoz 2007).

For a Jupiter-type planet at an orbital distance below 0.02 AU near a Sun-like star, the tidal forces and Roche lobe effects, which are not included in our model, become important (Garcia Munoz 2007, Penz et al. 2008, Erkaev et al. 2007). Therefore, we restrict the considered range of orbital distances by 0.02 AU. To justify the omission of tidal force a test calculation was performed for a tidally locked system with the gravity potential in equatorial plane $mG\frac{M_p}{r}\left(1+\frac{1}{2}(r/r_L)^3\right)$, where $r_L$ is a distance to the Lagrange point which for the considered HJ's (HD209458b) system is $r_L \approx 4.6$. It was found that the tidal force increases the escape velocity (on the day side of the planet). However, the mass loss rate increases only by 10% while maximum temperature decreases by only 10%. Therefore, it may be concluded that within the above made assumptions the main results of our study are not strongly affected by the inclusion of tidal force, though it should be taken into account for the completeness of the model of the HD209458b system (after all the major effects are properly studied and understood). At the same time, since the main goal of this paper is to build a comprehensive (1D) self-consistent model of a HJ's upper atmosphere escape for its further generalization to the case of a magnetized planet, without inclusion of tidal force the role of other material escape drivers and their mutual relation becomes more clear and easily to understand.

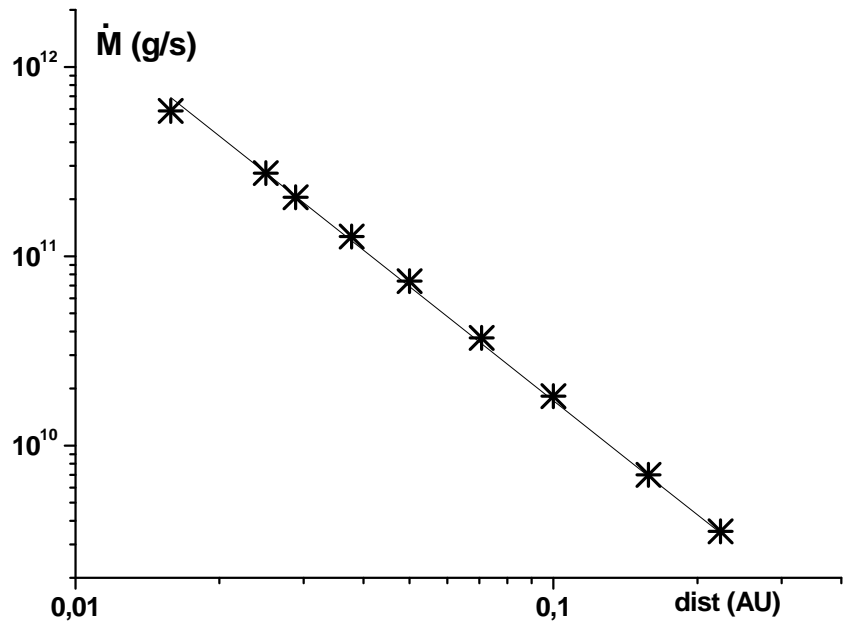

**Figure 4.** Dependence of mass loss rate on XUV flux expressed as an orbital distance of a planet to a Sun-like star. Straight line corresponds to $R^{-2}$.

*5.4 On the account of infrared cooling effect*

To study the role of infrared cooling effect in our computations, we followed the approach of Penz et al. (2008) and included the term $-n_a C_{ir}$ in the right hand side of equation (4). The taken coefficient $C_{ir} = 0.05\,\text{K/s}$ approximately corresponds to $H_3^+$ cooling rate reported in Yelle (2004) and García Muñoz (2007).

The $H_3^+$ is produced in reactions which involve ionized molecular hydrogen $H_2^+$ in the region where $H_2$ is a dominant specie instead of atomic hydrogen. To take this fact into account and restrict the infrared cooling by layers populated by molecular hydrogen we included reactions of molecular hydrogen dissociation and association: $H_2 + M \leftrightarrow 2H + M$ with the rates provided in

Yelle (2004) and García Muñoz (2007). Assuming local balance between $H_2$ and H as determined by rates of association and dissociation the relation between species is given by

$$n_{H2} = \frac{8 \cdot 10^{-33}(300/T)^{0.6}}{1.5 \cdot 10^{-9} e^{-48000/T}} \cdot n_H \cdot n_H \tag{19}$$

In the presence of $H_2$ the energy equation (4) should be modified. At short wavelengths the cross-section of XUV absorption by $H_2$ is about 2.5 larger than that for H. Without making a big error we can put it to be twice larger, -- that is the same cross-section as for H but for two hydrogen atoms comprising the hydrogen molecule. Finally, the heating term in eq. (4) may be modified corresponding to $n_H \sigma_{XUV,H} + n_{H2} \sigma_{XUV,H2} \approx (n_{H\Sigma} + n_{H2}) \cdot \sigma_{XUV,H}$. Therefore, in the infrared cooling term $-n_a C_{ir}$ instead of $n_a$ we use the sum density, $n_{H\Sigma} = n_H + n_{H2}$. After combining momentum equations for $H_2$ and H, the momentum equation can also be expressed in terms of the sum density $n_{H\Sigma}$ and average mass $\overline{m} = m \frac{n_H + 2n_{H2}}{n_H + n_{H2}}$:

$$\frac{\partial}{\partial t}\mathbf{V} + (\mathbf{V}\nabla)\mathbf{V} = -\frac{1}{\overline{m} \cdot n_{H\Sigma}} \nabla n_{H\Sigma} kT - \frac{1}{\overline{m} \cdot n_{H\Sigma}} \nabla n_e kT + \nabla \frac{GM_p}{r} \tag{20}$$

This equation replaces equation (3) in the considered set of model equations. By this, average mass $\overline{m}$ can be expressed through sum density $n_{H\Sigma}$ using the equation (19).

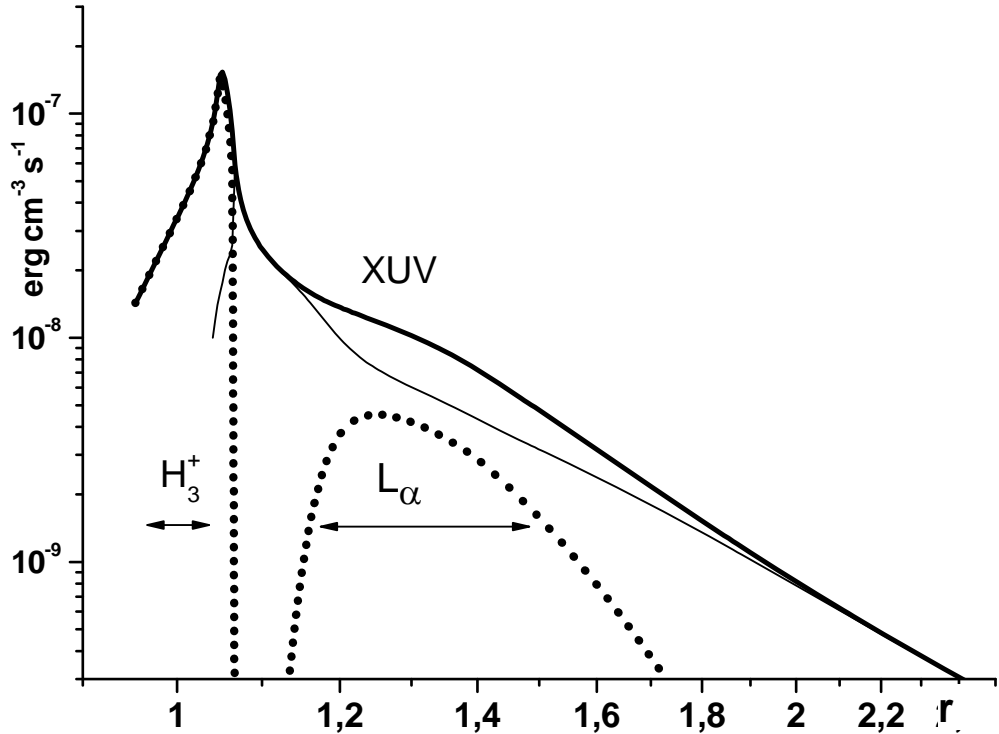

**Figure 5.** Spatial dependencies of XUV absorption rate (thick solid line), advection and adiabatic cooling rate (thin solid line) and radiative cooling rate due to infrared and Lyman-alpha emission (dotted line).

The corresponding energy density rates for XUV absorption, advection plus adiabatic, and radiation cooling obtained by solving the modified equations of the model are shown in Figure 5. With the inclusion of infrared cooling $C_{ir}$ the peak XUV absorption rate is about two times larger than in the calculation presented in in Figure 1. This happens because of deeper penetration of XUV due to generally smaller column density.

Note that radiative cooling consists of an infrared part at low temperatures and a Lyman-alpha part at sufficiently high temperatures. The maximum infrared cooling rate reaches maximum XUV heating rate, and both have value slightly larger than $10^{-7}$ erg·cm$^{-3}$·s$^{-1}$, which agrees well with the results of Yelle (2004). The value of pressure at this maximum is about $0.07\,\mu$bar. Transition from molecular to atomic hydrogen occurs at slightly larger height where a pressure of 0.015 $\mu$bar and temperature of about 2000 K are realized. One can see that close to the conventional planet surface, i.e. $r=1$ (in units of $R_p$) infrared cooling clearly dominates. Then, after some short space interval ($\sim 0.05 R_p$), Lyman-alpha radiation becomes to be an efficient coolant. Finally, at the heights above $1.5 R_p$ only the adiabatic and advection cooling counteract the XUV heating.

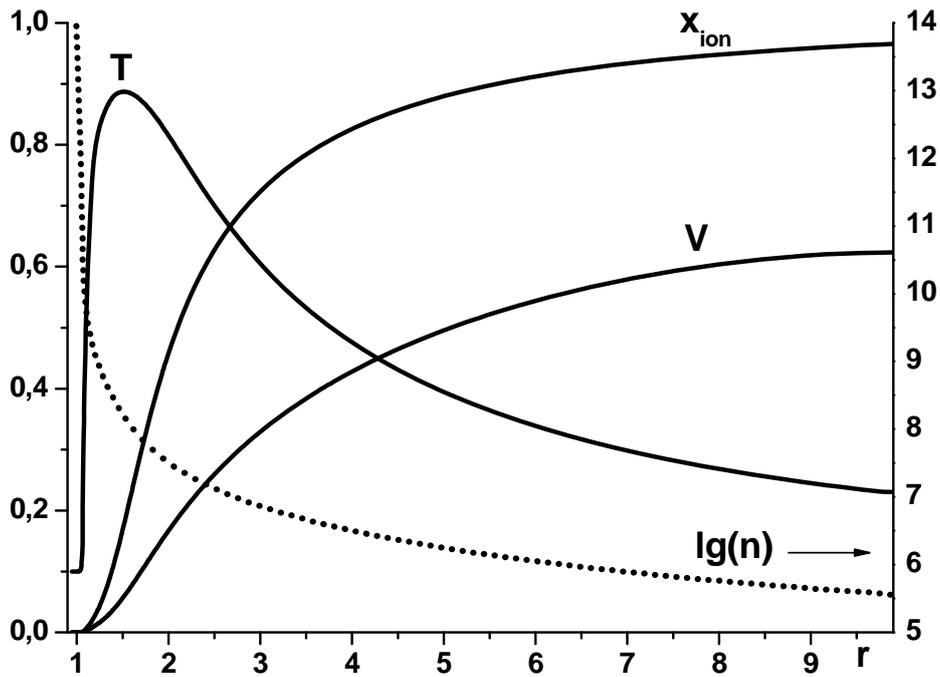

**Figure 6.** Profiles of temperature, velocity, ionization degree (left axis) and density (dotted line, right axis, log scale) in dimensionless units obtained in calculation with inclusion of H$_2$ chemistry and infrared cooling by $H_3^+$ for the model parameters from Table 1.

The spatial distribution profiles of main physical quantities obtained in course of calculations with the inclusion of the effects of H$_2$ chemistry and $H_3^+$ cooling are shown in Figure 6. Comparison of these results with those presented in Figure 1 reveals that account of the infrared cooling effects in the deep layers does not lead to any principal difference in the system behavior. Account of $H_3^+$ cooling effect as $\propto n_{H\Sigma} C_{ir}$ leads to the reduction of the material escape rate of about two times. A variation of the $C_{ir}$ value by an order of magnitude results only in a moderate (about a factor of 1.7) change in the mass loss rate. Thus, the effect of $H_3^+$ cooling has been estimated more or less robustly. This result differs from that of Penz et al. (2008), where no influence of infrared cooling was reported. In that respect we note that, in spite of the peak cooling rate in our case and in Penz et al. (2008) was the same, the XUV heating rate in Penz et al. (2008) is about an order of magnitude larger.

*5.5 Material ionization as a distributed acceleration factor*

Here we would like to demonstrate the performance of the above discussed (see subsection 3.1) effect of the "ionization boost" (due to the contribution of the electron and ion pressure gradients) in the increase the atmospheric material outflow speed.

It follows from the integration of equation (3) over r (from the planet surface to infinity) that the term with electron pressure gives a positive input in the kinetic energy of the expanding flow. The electron density increases with the radial distance over the integration interval from $n_e = 0$ at $r = R_p$ up to some maximum value and then again falls to zero, while the total number density of hydrogen $n$ monotonically decreases with distance. Due to that two areas can be distinguished depending on the radial distance r : 1) the area of increasing electron pressure (closer to the planet) where the pressure gradient acts as a decelerating factor on the plasma flow, and 2) the outer area of decreasing pressure where the electron pressure gradient accelerates the expanding plasma flow. Since the total density $n$ decreases exponentially fast with distance, and the location of maximum of electron pressure is spatially close to that of the maximum of temperature, the overall contribution of the electron pressure gradient term in the flow kinetic energy can be estimated as

$$-\int_{R_p}^{\infty} \frac{1}{n} \nabla n_e kT \approx kT_{max} \qquad (21)$$

The effect of acceleration of the escaping wind by the additional pressure gradient of the ionized gas is illustrated in Figure 7, where besides of the typical *free outflow* solution (the same as in Figure 1), the calculations without ionization and with zero electron pressure are shown. In the limiting case without ionization, the XUV radiation equally heats the whole volume of gas, resulting in a very large escape velocity and high temperature which continuously increase with distance. The speed of expanding wind exceeds the adiabatic sonic point $\sqrt{\gamma kT/m}$ at about $7.3R_p$. In another limiting case, shown in Figure 7, the effect of material ionization is included. This results in spatial dependence of XUV heating and its strong decrease at heights beyond several $R_p$. At the same time, the electron pressure in this limiting case (i.e. the ionization boost) is set to zero, so that the effect of additional pressure gradient of the ionized gas is artificially ignored. The resulting flow velocity in this case is essentially smaller than that in the true case and it never reaches the sound speed values, while the overall mass-loss is 1.5 times smaller. Contrary to these limiting cases, the true case, with the effects of ionization and the additional pressure gradient of the ionized gas (electron pressure) properly taken into account, is characterized by increase of outflow velocity which reaches a sonic point at the distance of $6.25R_p$. Note, that because of different spatial distribution of temperature in the cases considered here, the corresponding density profiles are also essentially different.

The comparison of the limiting cases presented in Figure 7 enables to conclude that the effect of ionization boost is essentially important in the case of a Hot Jupiter's expanding atmosphere. It appears a major mechanism that compensates to certain degree the decelerating action of adiabatic cooling and provides the necessary additional propulsion force to support the *free outflow* regime of atmosphere expansion.

Inclusion of the tidal force effects in the comparative study of the "ionization boost" yields that it slightly but not principally affects the results of the modelling. In particular while in the case without the tidal force the effect of the "ionization boost" at r=8 Rp causes the increase of the escape velocity by $\Delta V \approx 0.5$ (see Fig.7), the inclusion of the tidal force under the same other conditions results in $\Delta V \approx 0.4$.

It should be noted that electron pressure acts primarily on ions. In the outer, highly ionized, layers of the expanding atmosphere, where the electron pressure gradient acts as an accelerating force, the ions naturally appear as a major mass contributing component. As to the inner regions populated by partially ionized plasma, it was shown by Guo (2011) that due to fast charge exchange the added momentum of ions is effectively transferred to atoms.

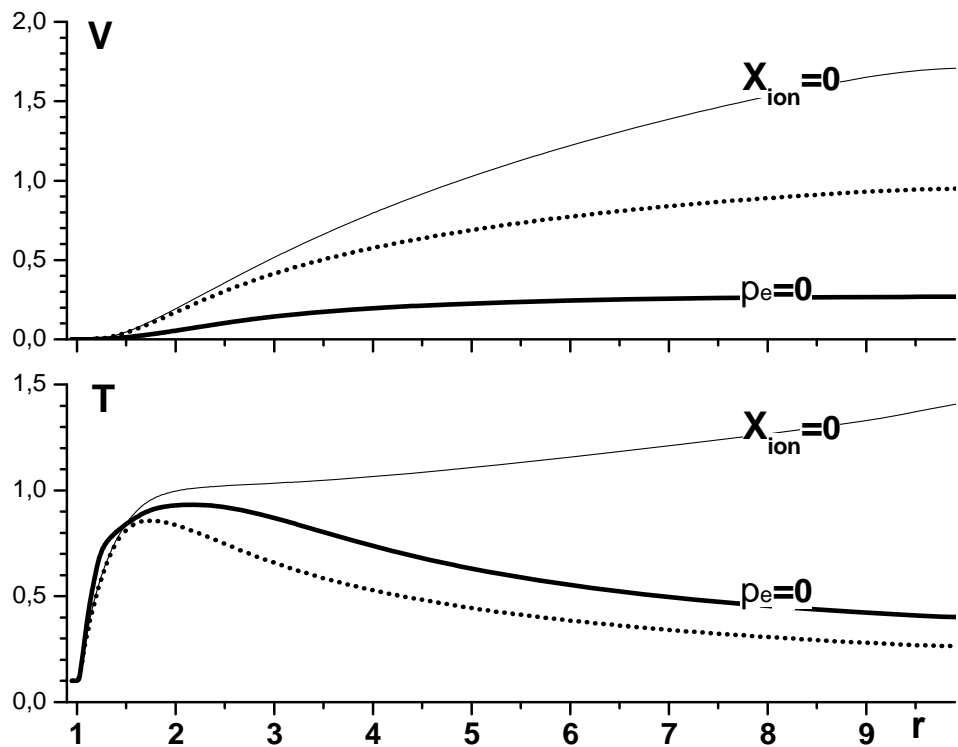

**Figure 7.** Profiles of velocity and temperature in true case (dotted lines), zero electron pressure case (thick lines) and no ionization case (thin lines).

*5.6 Restricted flow regime of a Hot Jupiter's atmosphere expansion*

When the *closed* condition is applied at the outer boundary of the simulation box, the expanding atmospheric material flow gradually decays until a steady state with an almost zero flux through the outer boundary is achieved. This situation corresponds to the *suppressed* or *restricted* flow regime. Because of the absence of the adiabatic cooling in this case, the energy balance condition is provided due to Lyman-alpha emission. This is true everywhere except of cold, non-ionized and dense thermosphere with a temperature of several thousand K and less. In reality, in the case of complete absence of material escape, a solution with a temperature at the inner boundary of the order of 1000 K, provided only by Lyman-alpha cooling, is not possible. Therefore, in our numerical model we allow a small outflow velocity at the outer boundary, so that the corresponding small flow in the dense thermosphere can sustain energy balance with the XUV heating. Under these circumstances, the material escape rate is proportional to the allowed fixed outflow velocity at the outer boundary. It is worth, in that respect, to compare different temperature profiles realized for different outflow velocities.

Figure 8 shows the temperature profiles obtained for different values of the outer boundary outflow velocity. With a series of consequent simulation runs we found that for $V_{out} < 0.1\,\text{km/s}$ the solution does not depend on the outflow velocity $V_{out}$ at the outer boundary anymore. By this, the maximum value of the velocity achieved inside the simulation box is about 2-3 times larger than the outer boundary outflow velocity, but still remains rather small $V_{max} < 0.25\,\text{km/s}$. At fixed outer boundary velocity $V_{out} = 0.1\,\text{km/s}$, the mass loss rate is 3.5 times smaller than in the case of *free outflow* (i.e. open outer boundary).

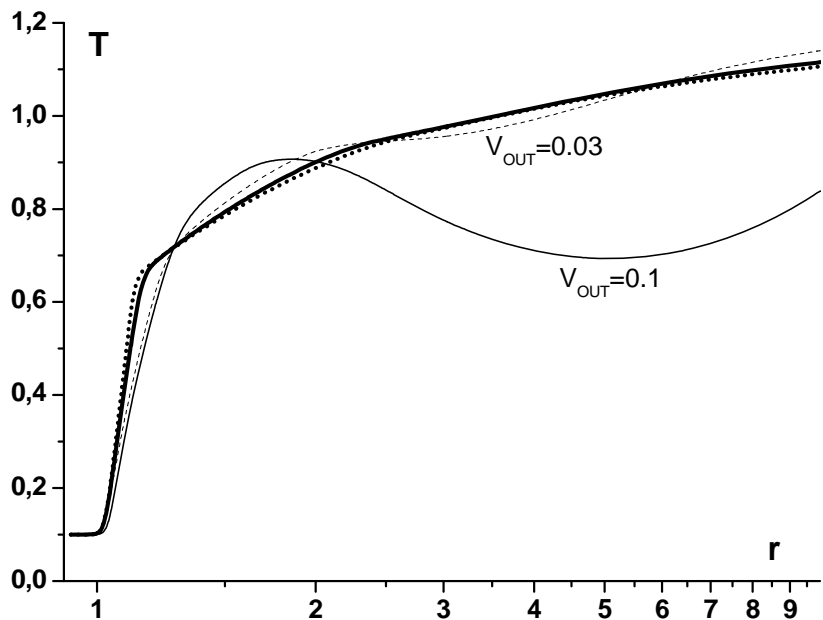

**Figure 8.** Temperature profile in the case of *restricted flow* regime for different fixed outflow velocities: 0.1 (thin solid line), 0.03 (dashed), 0.01 (thick solid) and 0.003 (dotted).

It can be seen in Figure 8 that the material temperature grows much faster near the planet surface as compared to the case of an *open* outer boundary condition. The temperature grows continuously over large distances, because the adiabatic cooling doesn't work in this case, except of only a restricted region at $r \leq 1.15 R_p$ close to the planet surface. The pressure averaged temperature (see equation (14)) in the interval of heights $R_p \leq r \leq 3 R_p$ is about 9000 K. This leads to the formation around a planet of the gas envelope which is several times (up to an order of magnitude) denser and somewhat hotter, as compared to the case of *free outflow* regime.

As already pointed out, the temperature in the case of a *restricted flow* regime depends on the efficiency of Lyman-alpha cooling. In view of the fact that deep near the planet surface the atmospheric gas is optically thick for Lyman-alpha, it is important to check also possible influence of Lyman-alpha re-absorption (included in the equation (11)). It becomes important when the number of Lyman-alpha photons is sufficiently large: $\tilde{N}_{L\alpha} e^{E21/T} \sim \tau_{21} \Lambda_e \cdot \sigma_{21} (n_a \sigma_{L\alpha} R_p)^2 > 1$.

This estimation follows from eq. (12) after approximation of spatial derivatives as $\nabla\nabla \sim R_p^{-2}$. For the conditions of *restricted flow* the re-absorption affects Lyman-alpha cooling rate at heights below $2R_p$ (see Figure 9). Figure 9 shows the result of model calculation with a relatively high Lyman-alpha absorption cross-section $\sigma_{L\alpha} \sim 10^{-14} T^{-1/2}$ cm$^2$.

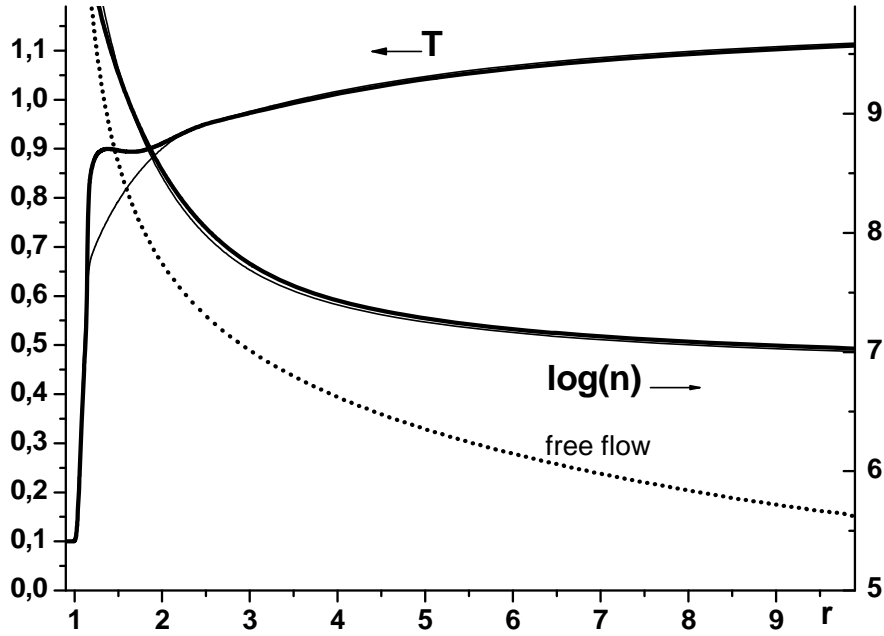

**Figure 9.** Temperature and density distributions in the case when Lyman-alpha re-absorption is included in the model (thick solid lines) and without it (thin lines). *Closed* condition (i.e., *restricted flow* regime) is applied at the outer boundary in both cases. For comparison, the dotted line shows density distribution in the case of *open* (i.e., *free outflow* regime) outer boundary condition.

As expected, because of strong re-absorption and established balance between hydrogen excitation and de-excitation which diminishes the efficiency of cooling, the temperature visibly increases at distances below $2R_p$. Maximum increase of temperature as compared to the case without Lyman-alpha re-absorption is about 1500 K. Farther from the planet the effect of Lyman-alpha re-absorption decreases and for $r > 2.5 R_p$ it becomes negligible. As to the influence of Lyman-alpha re-absorption on the gas density at distances up to several $R_p$, i.e. inside the dead zone region which might be formed in the presence of magnetic field, as it can be seen in Figure 9, such influence is rather small and practically undistinguishable in the numerical calculation result. Therefore, depending on purpose of particular numerical simulation, in some cases the effect of Lyman-alpha re-absorption can be neglected, despite of its obvious influence on the temperature of gas close to the planet.

Figure 9 allows also to compare the density distributions realized in the cases of *open* and *closed* outer boundary conditions, i.e. in the *free outflow* and *restricted flow* regimes, which may correspond, for example, the "wind-" and "dead-" zones of a Hot Jupiter magnetosphere (Trammell et al. 2011, Khodachenko et al. 2012). In particular, the plasma density in the case of *restricted flow* regime is up to an order of magnitude higher than in the case of *free outflow*.

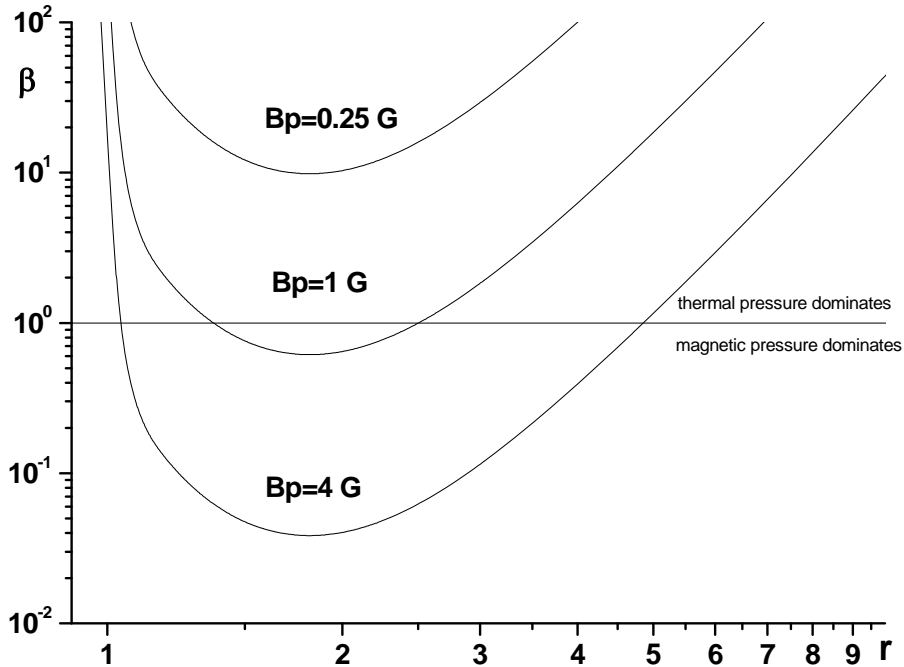

**Figure 10.** Spatial behavior of plasma thermal beta for different strengths of a magnetic field at the planet's surface.

Based on the obtained numerical solutions, it is possible to estimate strength of a Hot Jupiter's magnetic field, needed to affect and confine the hot gaseous envelope around the planet, energized by the stellar XUV. For this purpose we check spatial behavior of plasma beta $8\pi nT/B^2$ in the equatorial plane for different values of magnetic field at $r = R_p$, assuming for the definiteness sake an undisturbed magnetic dipole field geometry and a quasi-stationary *restricted flow* regime. The result is presented in Figure 10, which in particular shows, that the surface fields larger than 1 G should strongly affect the plasma outflow. The area of strongest magnetic control is located near $2R_p$, whereas the overall size of the magnetic field dominated region may reach, depending on the field strength, up to several $R_p$. Note that because of the steep density increase close to the planet surface, thermal pressure exceeds magnetic pressure even for very large surface fields.

Another interesting issue concerns the flow above the dipole poles. As it is well known, if plasma moves along the magnetic field lines in a strong dipole field, the divergence of material flux (guided by the magnetic field lines) at pole regions is larger than that in the case of the spherically isotropic expansion. Since the adiabatic cooling is crucial for sustaining the temperature of the expanding material, a certain reduction of the temperature and escape velocity in the polar regions, as compared to the isotropic expansion case, may be expected. This effect is demonstrated in Figure 11 which shows the outcome of our 1D HD simulation with an artificially increased ("cubic") divergence term as $\partial V/\partial r + 3V/r$ (analogous to divergence of dipolar magnetic field in polar regions), taken instead of the true divergence term $div\mathbf{V} = \partial V/\partial r + 2V/r$ which corresponds to an isotropic spherically symmetric expansion. Justification for such substitution based on the dipole field geometry can be found, for example, in Trammell et al. (2011). One can see in Figure 11 that the outflow velocity and density in the case of increased divergence are significantly smaller. By this, the maximum temperature realized in the case of inflated flow is by ~1000 K lower than that for the spherically symmetric isotropic flow. The sonic point is reached at about $8R_p$ and father than in equatorial case contrary to expectations.

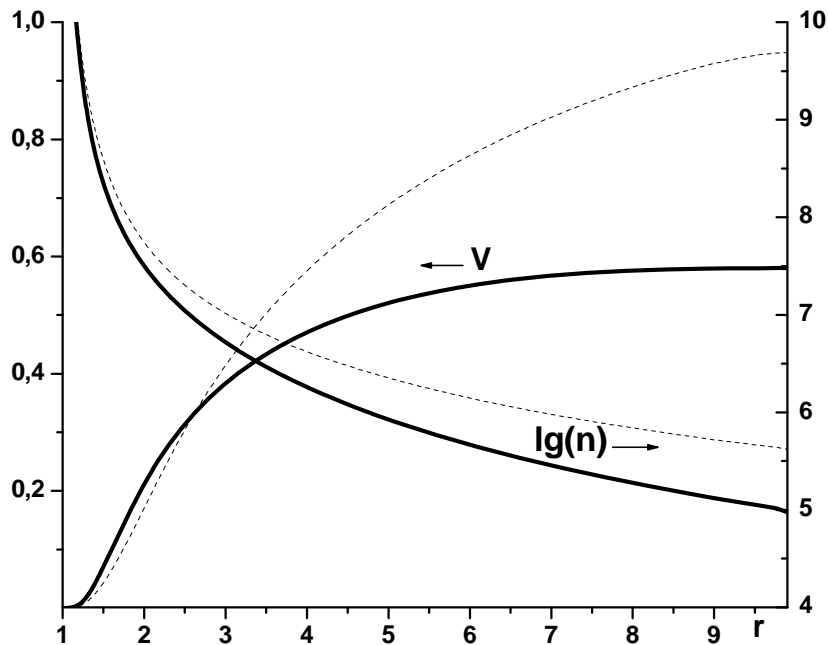

**Figure 11.** Velocity and density in case of an artificially increased ("cubic") divergence (solid lines) in comparison with the true divergence case (dashed lines).

## 6. Discussion and conclusions

We investigated the formation and escape regimes of a hot plasma envelope of a Jupiter-type exoplanet created due to XUV heating of the planetary hydrogen atmosphere with taking into account the processes of radiation transfer (i.e. excitation/de-excitation), ionization/recombination, and adiabatic cooling of the atmospheric gas. Our study confirmed the results reported in a number of previous works dedicated, as well as the present paper, mostly to the exoplanet HD209458b (Yelle 2004, García Muñoz 2007, Guo 2011, 2013, Koskinen et al. 2010, 2013) and revealed that a Hot Jupiter plasma envelope under the typical conditions of an orbital distance 0.05 AU around a Sun-type star should be rather hot, but with not very fast hydrodynamic escape resulting in the mass loss rates of about $(4-7) \cdot 10^{10}$ g/s. By this, the escaping material flow becomes supersonic at distances of about 10 planetary radii.

Note that, being interested in a pure thermal escape of a Hot Jupiter upper atmosphere caused only by the stellar radiation factors (i.e. material XUV heating and ionization) we did not include the effects of the stellar-planetary gravitational interaction. On the other hand, from the dedicated studies (Penz et al. 2008, Erkaev et al. 2007, 2011) it is known that in a system like that of

HD209458b the location of L1 point is at about $4,3R_p$. This means that the effects of the gravitational material escape due to the gas expansion beyond the Hill sphere add to the pure thermal mass loss, and should be properly included in a more complete model. At the same time, preliminary estimates provided in the present paper (see Figure 10) reveal that the intrinsic magnetic field may strongly influence the motion of the escaping planetary plasma up to the distances comparable with L1 in the HD209458b system. In this case the role of the gravitational escape effects will be strongly diminished whereas the dominating magnetic field will control the process of the mass loss.

To evaluate the physical drivers of the escape and the parameters of the created hot plasma envelope around the planet, we checked the influence of various processes such as infrared cooling by $H_3^+$ molecules in deep and cold layers of the planetary thermosphere, emission and re-absorption of Lyman-alpha photons, boundary and geometry conditions which may reflect different kinds of interaction with the planetary magnetic field.

Having in mind the primary goal of the presented in this paper 1D HD numerical modeling, as a first step in our *two-step modeling* approach aimed at understanding first of hydrodynamic behavior of the expanding non magnetized upper atmosphere of a giant exoplanet, significant attention was paid to the investigation of different types of boundary conditions and their influence on the achieved material escape regimes. Based on this understanding the next step (MHD modelling) with the consequent inclusion of the planetary magnetic field and relevant effects can be done in a more comprehensive and justified way. Moreover, several general conclusions regarding the influence of magnetic field on the plasma distributions, flows and temperature regimes were possible already in the present HD simulations, based on the results obtained for different types of inner and outer boundary conditions.

It was found that due to ability of the short wavelength photons to penetrate deeper in the atmosphere, the high energy part of the XUV spectrum, as a heater of the deep and dense layers of thermosphere, plays an important role in formation of a hot plasma envelope around the planet. This is directly related to the value of maximum pressure implied at the inner boundary of the planet. Using the proxy model of spectrally integrated solar XUV flux we derived a simple analytical estimate for the maximum pressure (see equation (18)) up to which XUV heating can be balanced by adiabatic cooling for the given base temperature of the atmosphere and mass loss rate. For the typical parameters of HD209458b this pressure is $\sim 10\,\mu bar$, i.e. higher than that used in previous models of the planet. Note, that the velocity of escaping gas at the height where this pressure is realized in the atmosphere is negligibly small as compared to the thermal speed, and the XUV spectral cut off is around the wavelength of about 2,5 nm. It was also demonstrated with a series of test simulations, that the inner boundary of the simulation box located at the heights of lower pressures produces proportionally smaller escape rates, whereas the boundary at larger pressures gives approximately the same results.

At the pressure values used at the inner boundary, the hydrogen is present mostly in a molecular form, and such species as $H_3^+$ may provide an efficient cooling effect by infrared radiation. By incorporation of these factors into the model we found that hydrogen chemistry at the base of the thermosphere might change the parameters of the formed hot extended plasma envelope and result in reduction of the material escape rate by a factor of about two.

In the present work we also made a deeper insight at the mechanism of formation of the escaping planetary wind. As it is known from the stellar wind studies, a supersonic flow of escaping material has to be supported by a mechanism of additional heating/energizing, which operates in the distance range where the flow gains its most velocity. Various mechanisms of wave heating are traditionally referred for the stellar winds in that regard, though the exact process remains unknown. In the case of a Hot Jupiter, it appears that the XUV heating of atmospheric gas alone is not enough to balance the adiabatic cooling at distances of several planetary radii, because gas becomes ionized and the radiation absorption decreases at heights $>3R_p$. However, the same effect of gas ionization, which reduces the efficiency of XUV heating, plays an important role of a booster of the escaping wind due to the buildup of an additional electron pressure gradient force. The thermal pressure almost doubles at the ionization front due to appearance of additional electron pressure which contributes the acceleration of the escaping partially ionized plasma flow. This effect means that part of absorbed stellar XUV energy is stored in thermal energy of electrons which then is converted into kinetic energy of the expanding plasma flow. In that respect, self-consistent account of the processes of gas ionization and recombination is crucial for the correct calculation of planetary atmospheric material escape. Simple use of the prescribed or fixed ionization ratios may lead to unrealistic and erroneous results in that respect.

We investigated two types of HD solutions which represent two basic regimes (*restricted flow* and *free outflow*) of the expanding gaseous envelope of a giant exoplanet heated by stellar XUV. In fact, both regimes can be realized in the same physical system depending only on type of the outer boundary conditions (*closed* or *open*), which may in their turn be a consequence of specific action of the planetary magnetic field.

In the *free outflow* regime the adiabatic cooling alone provides efficient balance to XUV heating. It dominates over the radiation cooling everywhere except of the base of dense and cold thermosphere where infrared radiation is more efficient. Lyman-alpha radiation cooling may compete with the adiabatic cooling only in a thin layer around temperature maximum and as we have found out, only if the mass loss rate is below $5 \cdot 10^{10}$ g/s. It was found that at orbital distances from (0.25 - 0.1) AU where the effects of stellar gravitation may be still unimportant, the mass loss rate is mainly defined by the expansion and escape of the hot planetary atmospheric material and it is proportional to the stellar XUV flux.

The kinetic $mV^2/2$ and thermal $3kT/2$ energies per particle in a fully ionized escaping gas are an order of magnitude smaller than the stored ionization energy $E_{ion}$. A comparable with $E_{ion}$ part in the energy budget is the energy of particles ($E_g$ = 8.6 eV) which exit the gravitation bounding of the planet. Since the energy deposed in the system by XUV flux illuminating the planet, is removed mainly by the expanding gas outflow (except of inessential part lost via Lyman-alpha photons), and this was confirmed to be true by the performed numerical calculations, the atmospheric mass loss can be estimated as $\dot{M} \approx 4\pi R_p^2 m \eta_{net} F_{XUV} / (E_{ion} + E_g) \approx 6 \cdot 10^{10}$ g/s. This value for $\eta_{net} = 50\%$ is just about 14% less than $\dot{M}$ obtained in course of numerical simulations. Note that heating efficiency $\eta_h$ is not included in these estimations, since the energy spent of the ionization of neutral atmospheric gas constitutes the major portion of the whole absorbed energy. This estimate indicates that in the range of considered stellar-planetary parameters and XUV fluxes the mass loss is indeed energy limited, as it was found in a number of other models (Lammer et al. 2003, Erkaev et al. 2007).

In the *restricted flow* regime, the XUV heating is balanced by gas excitation and radiation of infrared and Lyman-alpha photons. However, the infrared and Lyman-alpha cooling acts only at certain restricted not overlaping temperature ranges which correspond to particular regions in the atmosphere. As result, the layers with temperatures between 2000 K and 7000 K cannot be effectively cooled by radiation. Due to that a flow with a relatively small, of about $200-300$ m/s, velocities at the outer boundary is built up, that provides necessary additional adiabatic cooling. Mass conservation requires that the material is either removed through the outer boundary, or is settled in a kind of zonal convection.

Another specific detail of the *restricted flow* solution which we investigated is Lyman-alpha re-absorption. For the cross-section of about $10^{-15} - 10^{-14}$ $cm^2$ the atmospheric gas of the considered giant planet HD209458b is everywhere optically thick for Lyman-alpha photons. By this, the re-absorption leads to increase of population of excited hydrogen atoms and results in the decrease of Lyman-alpha cooling. The numerical solution which treats Lyman-alpha radiation transfer in diffusion approximation shows that optical thickness of the gaseous envelope leads to the increase of temperature by about 1000 K at distances below $2R_p$. At the same time, at larger distances the temperature and density remain nearly the same as in the case without inclusion of Lyman-alpha re-absorption.

Finally, based on the simulated escaping atmosphere of HD209458b, we find that the intrinsic planetary magnetic field which may have at the planetary surface a value higher than 1 G, may certainly affect the material flow and result in formation of the "dead" zone of stagnated plasma in the equatorial plane around the planet. As to the magnetic polar regions, the material escape rate there might be severely diminished due to faster than the spherically isotropic spatial divergence of the dipole magnetic field lines resulting in a stronger adiabatic cooling.

**Acknowledgements**

This work was supported by SB RAS Research Program grant II.8.1.4. RFBR grants 09-08-00970, 12-02-00367, 14-29-06036 and OFN RAS Research Program №22. MLK, HL, and KGK acknowledge the projects P25587-N27, S11606-N16, and S11607-N16 of the Austrian Science Foundation (FWF). The authors are thankful to EU FP7 project IMPEx (No.262863) for support of numerical modeling work and providing collaborative environment for research and communication. The authors also thank the International Space Science Institute (ISSI) in Bern, and the ISSI team "Characterizing stellar- and exoplanetary environments".

# Figure Captions

**Figure 1.**
Profile of temperature, velocity, ionization degree (left axis) and density (dotted line, right axis, log scale) in dimensionless units obtained in calculation with typical model parameters from Table 1.

**Figure 2.**
Temperature and velocity in dimensionless units as functions of pressure at the inner boundary in the thermosphere. Dotted line – analytic solution defined with equations (16) and (17).

**Figure 3.**
Mass loss rate as a function of the inner boundary pressure $\widetilde{P}_0$.

**Figure 4.**
Dependence of mass loss rate on XUV flux expressed as an orbital distance of a planet to a Sun-like star. Straight line corresponds to $R^{-2}$.

**Figure 5.** Spatial dependencies of XUV absorption rate (thick solid line), advection and adiabatic cooling rate (thin solid line) and radiative cooling rate due to infrared and Lyman-alpha emission (dotted line).

**Figure 6.**
Profiles of temperature, velocity, ionization degree (left axis) and density (dotted line, right axis, log scale) in dimensionless units obtained in calculation with inclusion of $H_2$ chemistry and infrared cooling by $H_3^+$ for the model parameters from Table 1.

**Figure 7.** Profiles of velocity and temperature in true case (dotted lines), zero electron pressure case (thick lines) and no ionization case (thin lines).

**Figure 8.** Temperature profile in the case of restricted flow regime for different fixed outflow velocities: 0.1 (thin solid line), 0.03 (dashed), 0.01 (thick solid) and 0.003 (dotted).

**Figure 9.** Temperature and density distributions in the case when Lyman-alpha re-absorption is included in the model (thick solid lines) and without it (dashed lines). Closed condition (i.e., *restricted flow* regime) is applied at the outer boundary in both cases. For comparison, the dotted line shows density distribution in the case of open (i.e., *free outflow* regime) outer boundary condition.

**Figure 10.** Spatial behavior of plasma thermal beta for different strengths of magnetic field at the planet's surface.

**Figure 11.** Velocity and density in case of an artificially increased ("cubic") divergence (solid lines) in comparison with the true divergence case (dashed lines).